\documentclass[twocolumn]{IEEEtran}
\usepackage[noadjust]{cite}
\usepackage[cmex10]{amsmath}
\usepackage{amsfonts}
\usepackage{amssymb}
\usepackage{algorithm}
\usepackage{algorithmic}
\usepackage[dvipsnames]{xcolor}
\usepackage[colorlinks,linkcolor=black,anchorcolor=black,citecolor=black]{hyperref}
\usepackage{graphicx}
\usepackage{verbatim}
\usepackage{indentfirst}
\usepackage{epsfig}
\usepackage{bm}
\usepackage{multirow}
\usepackage{subfigure}
\usepackage{setspace}
\usepackage{multicol}

\IEEEoverridecommandlockouts

\newcommand{\mv}[1]{\mbox{\boldmath{$ #1 $}}}

\begin{document}
\title{Integrating Intelligent Reflecting Surface into Base Station: Architecture, Channel Model, and Passive Reflection Design}
\author{Yuwei Huang, {\it Student Member, IEEE}, Lipeng Zhu, {\it Member, IEEE}, and Rui Zhang, {\it Fellow, IEEE}

\thanks{
Y. Huang is with the Department of Electrical and Computer Engineering, National University of Singapore, Singapore 117583, and also with the NUS Graduate School, National University of Singapore, Singapore 119077 (e-mail: yuweihuang@u.nus.edu).

L. Zhu is with the Department of Electrical and Computer Engineering, National University of Singapore, Singapore 117583 (e-mail:zhulp@nus.edu.sg).

R. Zhang is with School of Science and Engineering, Shenzhen Research Institute of Big Data, The Chinese University of Hong Kong, Shenzhen, Guangdong 518172, China (e-mail: rzhang@cuhk.edu.cn). He is also with the Department of Electrical and Computer Engineering, National University of Singapore, Singapore 117583 (e-mail: elezhang@nus.edu.sg).
}}
\maketitle
\begin{abstract}
Intelligent reflecting surface (IRS) has emerged as a cost-efficient technique to improve the wireless network's capacity and performance. Existing works on IRS have mainly considered IRS being deployed in the environment to dynamically control the wireless channels between the base station (BS) and its served users in favor of their communications. In contrast, we propose in this paper a new integrated IRS-BS architecture by deploying IRSs inside the BS's antenna radome to directly reconfigure the signal radiation to/from the BS's antennas. In other words, the IRSs can be considered as auxiliary passive arrays with real-time reconfigurability equipped at the BS to enhance its communication performance cost-effectively. Since the distance between the integrated IRSs and BS's antenna array is practically small (in the order of several to tens of wavelengths), the path loss among them is significantly reduced as compared to conventional IRS deployed much farther away from the BS, while the real-time control of the IRS's reflection by the BS becomes easier to implement. However, the resultant near-field channel model also becomes drastically different from its far-field counterpart for conventional far-away IRSs in the literature. Thus, we propose an element-wise channel model for IRS to characterize the channel vector between each single-antenna user and the antenna array of the BS, which includes the direct (without any IRS's reflection) as well as the single and double IRS-reflection channel components. Based on this channel model, we formulate a problem to optimize the reflection coefficients of all IRS reflecting elements for maximizing the uplink sum-rate of the users.  By considering two typical cases with/without perfect channel state information (CSI) at the BS, the formulated problem is solved efficiently by adopting the successive refinement method and iterative random phase algorithm (IRPA), respectively.  Numerical results validate the substantial capacity gain of the integrated IRS-BS architecture over the conventional multi-antenna BS without integrated IRS. Moreover, the proposed algorithms significantly outperform other benchmark schemes in terms of sum-rate, and the IRPA without CSI can approach the performance upper bound with perfect CSI as the training overhead increases.
\end{abstract}
\begin{IEEEkeywords}
Intelligent reflecting surface (IRS), integrated IRS-BS architecture, element-wise channel model, passive reflection design.
\end{IEEEkeywords}

\section{Introduction}
\IEEEPARstart{W}{ith} the development of digitally-controlled metamaterial technologies, intelligent reflecting surface (IRS) has emerged as a cost-efficient technique to achieve smart and reconfigurable radio environment for future wireless communication systems. Specifically, IRS is a programmable metasurface consisting of a large number of passive reflecting elements, whose amplitude and/or phase shifts can be individually controlled in real time, thereby enabling dynamic control over the wireless propagation channel for a variety of purposes (e.g., passive beamforming, multi-path diversity, interference nulling/cancellation, and so on) \cite{tutorial,magazine,tutorial_new}. Moreover, IRS dispenses with transmit and receive radio frequency (RF) chains and reflects ambient signals only passively, which requires low hardware cost and energy consumption. Thus, IRS has been extensively investigated in assorted wireless communication systems, such as orthogonal frequency division multiplexing (OFDM)-based wideband communications \cite{ofdm,ofdm1}, multi-antenna and/or multi-user communications \cite{antenna,antenna1}, multi-cell networks \cite{multi_cell,cell1}, non-orthogonal multiple access (NOMA) systems \cite{noma,noma1}, relaying communications \cite{relay,relay1}, physical-layer security \cite{security,security1}, unmanned aerial vehicle (UAV) communications \cite{uav,uav1}, mobile edge computing (MEC) \cite{MEC,MEC1}, etc.

To fully reap their benefits, IRSs need to be properly deployed between the base station (BS) (or access point (AP)) and its served users to decrease the product-distance path loss of the BS-IRS-user cascaded channels \cite{tutorial}. Towards this end, a promising solution is to deploy the IRS closer to the user terminals, e.g., in hotpot area, cell edge, or vehicles. Following this approach, the authors in \cite{ofdm} considered a single-user OFDM communication system by deploying an IRS in the vicinity of a single-antenna user, where the passive reflection of the IRS was optimized to maximize the achievable rate. In \cite{array}, a user-side IRS was employed to assist in the communication between a multi-antenna BS and a multi-antenna user, where the capacity of the considered multiple-input multiple-output (MIMO) system was maximized by alternately optimizing the passive reflection of the IRS and the covariance matrix of the transmit signals. Besides, the authors in \cite{moving} employed an IRS in a high-speed vehicle to enhance the communication performance between the users on board and a remote BS, where an efficient two-stage transmission protocol was proposed to conduct channel estimation and IRS reflection optimization. In addition to the user-side deployment, IRSs can also be deployed in the vicinity of the BS to assist in its communication with users distributed at different locations. In this regard, the authors in \cite{learning1} and \cite{learning2} deployed one IRS near the BS and adopted the deep reinforcement learning (DRL) technique to design the passive reflection of the IRS for assisting the communication between the BS and multiple users. Moreover, the authors in \cite{my} proposed an IRS-empowered BS architecture by deploying multiple IRSs close to the BS (e.g., in the range of several to tens of meters), where a novel user-IRS association design was proposed to select a certain number of  cascaded user-IRS-BS channels for estimation under a pre-determined channel training overhead, and then the reflection coefficients of all IRSs were jointly optimized to maximize the minimum achievable rate among the users based on the estimated channel state information (CSI). Furthermore, a hybrid deployment strategy was proposed in \cite{deploy} to combine the complementary advantages of the BS- and user-side IRSs, which can yield higher design flexibility and provide additional passive beamforming gain by exploiting the inter-IRS reflections. In this context, the authors in \cite{double2} investigated a double-IRS system with one IRS deployed near the multi-antenna BS and the other IRS deployed closer to the multi-antenna user, where all the involved channels were assumed to have dominant line-of-sight (LoS) paths, and the reflection coefficients of the two IRSs were jointly optimized to maximize the capacity of the MIMO system. Subsequently, the authors in \cite{double1} extended the double-IRS system to a more practical scenario with multiple users, where an efficient algorithm by exploiting alternating optimization was proposed for cooperative IRSs' beamforming design to maximize the minimum signal-to-interference-plus-noise ratio (SINR) among all users.

Note that both user-side and BS-side deployments in the aforementioned works can be categorized into the standalone IRS, where the IRSs are deployed in the environment and thus sufficiently far away (i.e., more than hundreds of carrier wavelengths) from the users/BSs. In practice, such deployment strategies still result in non-negligible product-distance path loss of the BS-IRS-user cascaded channels, which may limit the performance gain provided by IRS. Besides, to achieve full passive beamforming gain, the IRSs should be connected to BSs via wireless/wired links to execute reflection control for both channel estimation and data transmission, which requires a prohibitive signaling overhead between the BS and IRSs, especially for the case of large-size IRS and/or large number of IRSs.

\begin{figure}
\centering
\includegraphics[width=8cm]{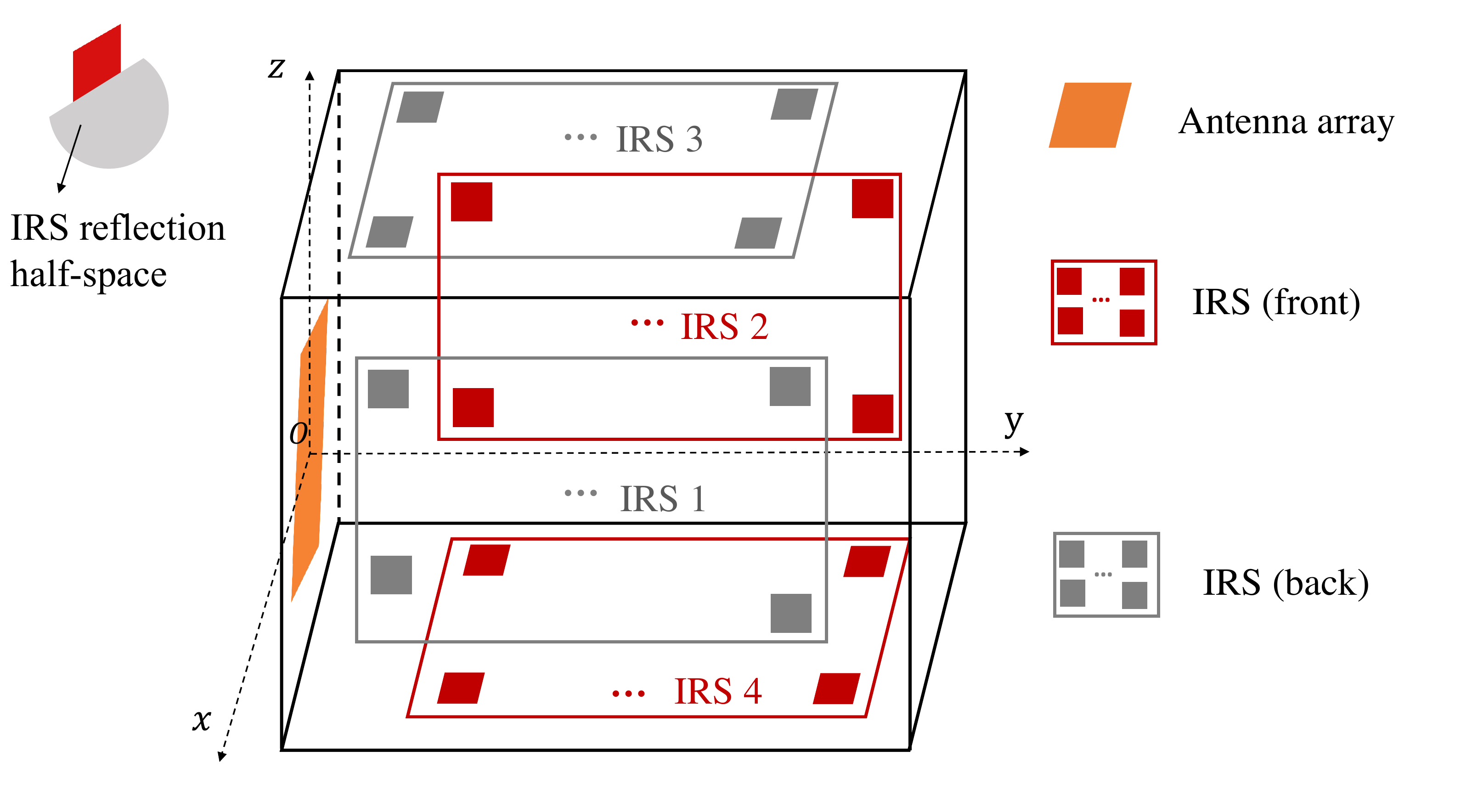}
\vspace{-10pt}\caption{Illustration of an integrated IRS-BS architecture.}\label{proposed_architecture}
\vspace{-10pt}\end{figure} 
To circumvent the above difficulties, in this paper we propose a new {\it integrated IRS-BS architecture}, where four IRSs with different orientations and an antenna array are deployed within the same cuboid antenna radome at the BS, as shown in Fig. \ref{proposed_architecture}. Due to the ultra-short distance between the IRSs and BS's antenna array (i.e., in the order of several to tens of wavelengths), the path loss of the reflection channels via IRSs can be significantly reduced. In addition, besides the single-IRS reflections between the users and the antenna array of the BS, the double reflections among the IRSs become much stronger than their counterparts in the conventional double-IRS systems with IRSs deployed much farther away from each other \cite{double1,double2}, and thus their effect to the system performance becomes significant and may even dominate over the single-IRS reflections. Moreover, the signaling overhead between the BS and IRSs is generally reduced because of their integrated architecture. Despite the above mentioned advantages, new technical issues also arise, which need to be investigated for our proposed integrated IRS-BS architecture. First, the conventional uniform plane wave (UPW) assumption cannot be directly applied to model the channels between the BS antenna array and each of IRSs as a whole since the far-field condition is not satisfied between them \cite{near_field1,near_field2,near_filed_zy}, which motivates us to develop a more practical and sophisticated channel model for the proposed integrated IRS-BS system. Second, the signals from/to the users  experience single and double reflections by multiple integrated IRSs, and the coupling effect of these channel components makes the optimization of IRS passive reflection more intractable as compared to the conventional design considering single-IRS reflections only. Third, the CSI becomes more substantial and is harder to obtain in practice due to the additional inter-IRS reflections involved, despite various channel estimation methods have been proposed for single-IRS \cite{group,liuliang,anchor,sparsity} or double-IRS systems \cite{double_estimation} (see \cite{survey,channel_estimation_survey}, and the reference therein). This is because our proposed architecture has multiple IRSs integrated very close to the BS's antenna array, and thus the number of channel coefficients to be estimated is much larger than that in conventional single-/double-IRS systems, which may require extremely high pilot overhead for estimating all reflection channels via all IRSs. To tackle the above challenges, we develop efficient algorithms for designing the IRS reflection coefficients under the case with/without perfect CSI at the BS, respectively. The main contributions of this paper are summarized as follows.
\begin{itemize}
\item First, considering the ultra-short distance between the IRSs and BS's antenna array, we propose an element-wise channel model for IRS to characterize the channel vector between each single-antenna user and the antenna array of the BS, which consists of the direct (without any IRS's reflection) as well as the single and double IRS-reflection channel components. Specifically, the UPW propagation is utilized to model the channels from each reflecting element (instead of the whole surface of the IRS) to each antenna at the BS as well as between the reflecting elements of different IRSs, where the reflection gain of each IRS reflecting element considers the non-isotropic electromagnetic response of signals incident from/reflected to different directions. In addition, to integrate more reflecting elements into the antenna radome, we generalize the integrated IRS-BS architecture by employing modular antenna arrays, where the whole antenna array is divided into several antenna modules, and each antenna module is surrounded by four smaller-size IRSs.
\item Next, we formulate an optimization problem to maximize the uplink sum-rate of the users by designing the reflection coefficients of all IRSs, which is non-convex and thus difficult to solve. To tackle this problem, we consider two typical cases with/without perfect CSI at the BS and develop efficient algorithms for solving the problems in both cases. For the case with perfect CSI, we adopt the successive refinement method to alternately update the reflection coefficient of each reflecting element by fixing those of other reflecting elements. While for the case without CSI, we extend the conventional random phase algorithm (RPA) to the iterative random phase algorithm (IRPA) for optimizing the reflection coefficients of multiple IRSs in an iterative manner. Specifically, we first decouple the reflection coefficients of multiple IRSs, and then adopt the RPA to update the reflection coefficients of each IRS by fixing those of other IRSs.
\item Finally, numerical results are provided to validate the efficacy of the proposed integrated IRS-BS architecture over the conventional multi-antenna BS without integrated IRS. The proposed algorithms for IRS passive reflection design are shown to achieve considerable performance gain as compared to other benchmark schemes in terms of sum-rate, and the proposed IRPA without perfect CSI can approach the performance upper bound with perfect CSI as the training overhead increases. Besides, the results demonstrate that the double-reflection channel components have a more dominant impact on the achievable sum-rate than the single-reflection channel components. Moreover, the effects of key system parameters, including the total number of reflecting elements, the suspension angle of the BS antenna radome, and the number of antenna modules in the generalized IRS-BS architecture, are also evaluated. The results show that the proposed integrated IRS-BS architecture is a cost-efficient design for future-generation BSs/APs in wireless networks, since the passive IRSs deployed in the near-field of the BS antenna array can serve as auxiliary passive arrays that can be configured jointly with the antenna array to significantly enhance the degrees of freedom (DoFs) for dynamically controlling the signal radiation to/from the BS.    
\end{itemize}

It is worth noting that there have been other architectures proposed for deploying IRSs at the BSs (or APs), such as active holographic MIMO surfaces \cite{hMIMO1,hMIMO2}, dynamic metasurface antennas (DMAs) \cite{dma1,dma2,dma3,dma4}, receiving IRSs \cite{receiving_irs}, and deploying IRSs for sensing \cite{sensing}. However, these architectures require to embed RF chains and signal processing units on IRS to control its analog beampattern for transmission and reception. In contrast, our proposed integrated IRS-BS architecture only attaches passive reflecting elements in the antenna radome, which can reconfigure electromagnetic propagation environments to enhance the communication performance without changing the RF front-end structure of existing BS/AP antennas. Thus, the proposed integrated IRS-BS architecture in general requires lower energy consumption and hardware cost, which is more compatible with the BSs/APs in existing wireless communication systems.

The rest of this paper is organized as follows. Section II presents our proposed system architecture and channel model. Section III presents the problem formulation and IRS passive reflection design with/without perfect CSI. Section IV provides numerical results to verify the efficacy of our proposed integrated IRS-BS architecture and  algorithms for IRS passive reflection design. Finally, Section V concludes this paper.

{\it Notation:} In this paper, scalars, vectors, and matrices are denoted by italic, bold-face lower-case, and bold-face upper-case letters, respectively. For a vector $\mv a$, $[\mv a]_{n}$ denotes its $n$-th entry. For a matrix $\mv A$, its transpose, conjugate transpose, and determinant are denoted as $\mv A^{T}$, $\mv A^{H}$, and $\det(\mv A)$, respectively. $\mv I_{M}$ denotes the identity matrix of size $M$. $\mathbb{C}^{x\times y}$ denotes the set of $x\times y$-dimensional complex-valued matrices, and $\mathbb{R}^{x\times y}$ denotes the set of $x\times y$-dimensional real-valued matrices. For a complex number $s$, $s^{*}$ and $|s|$ denote its conjugate and amplitude, respectively. $s\sim\mathcal{CN}(0,\sigma^{2})$ means that $s$ is a circularly symmetric complex Gaussian (CSCG) random variable with mean zero and variance $\sigma^{2}$. $i$ denotes the imaginary unit, i.e., $i=\sqrt{-1}$. For a vector $\mv a$, $\text{diag}(\mv a)$ denotes a diagonal matrix whose diagonal elements are specified by $\mv a$, and $\lVert \mv a \rVert$ denotes its 2-norm. Notation $\log_{2}(\cdot)$ denotes the logarithm function with base $2$; $\otimes$ represents the Kronecker product; and $\mathcal{O}(\cdot)$ denotes the Landau's symbol to describe the order of complexity.

\section{System Architecture and Channel Model}\label{system_architecture}
\subsection{System Architecture}
\begin{figure}
\centering
\includegraphics[width=8cm]{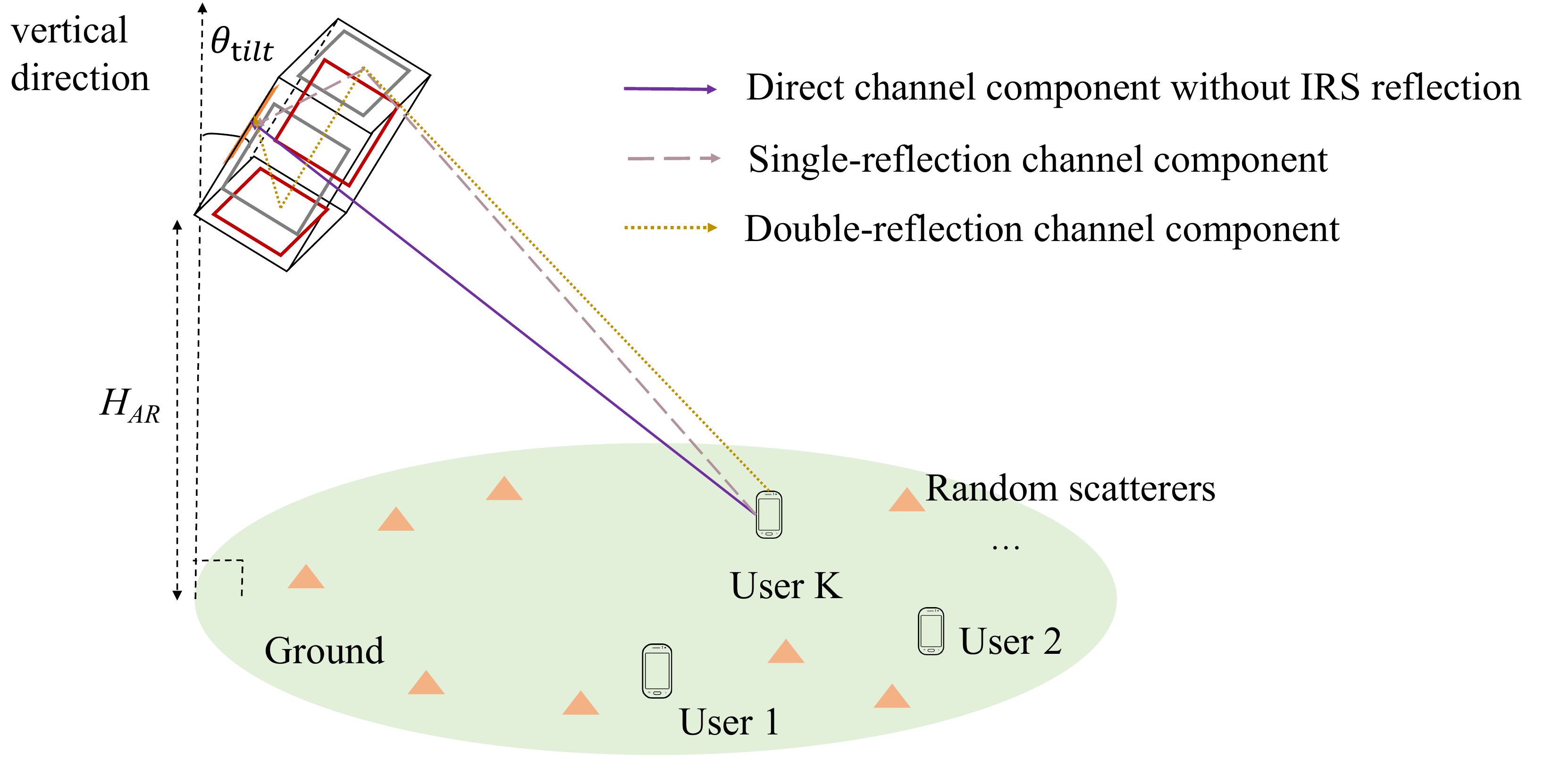}
\vspace{-10pt}\caption{System model of the integrated IRS-BS and its served users.}\label{system_model}
\vspace{-10pt}\end{figure}
\begin{figure} 
	\centering
	\subfigure[Cellular BS with $0\leq \theta_{tilt}\leq\frac{\pi}{12}$.]{
\begin{minipage}[t]{0.5\textwidth}
		\includegraphics[width=8cm]{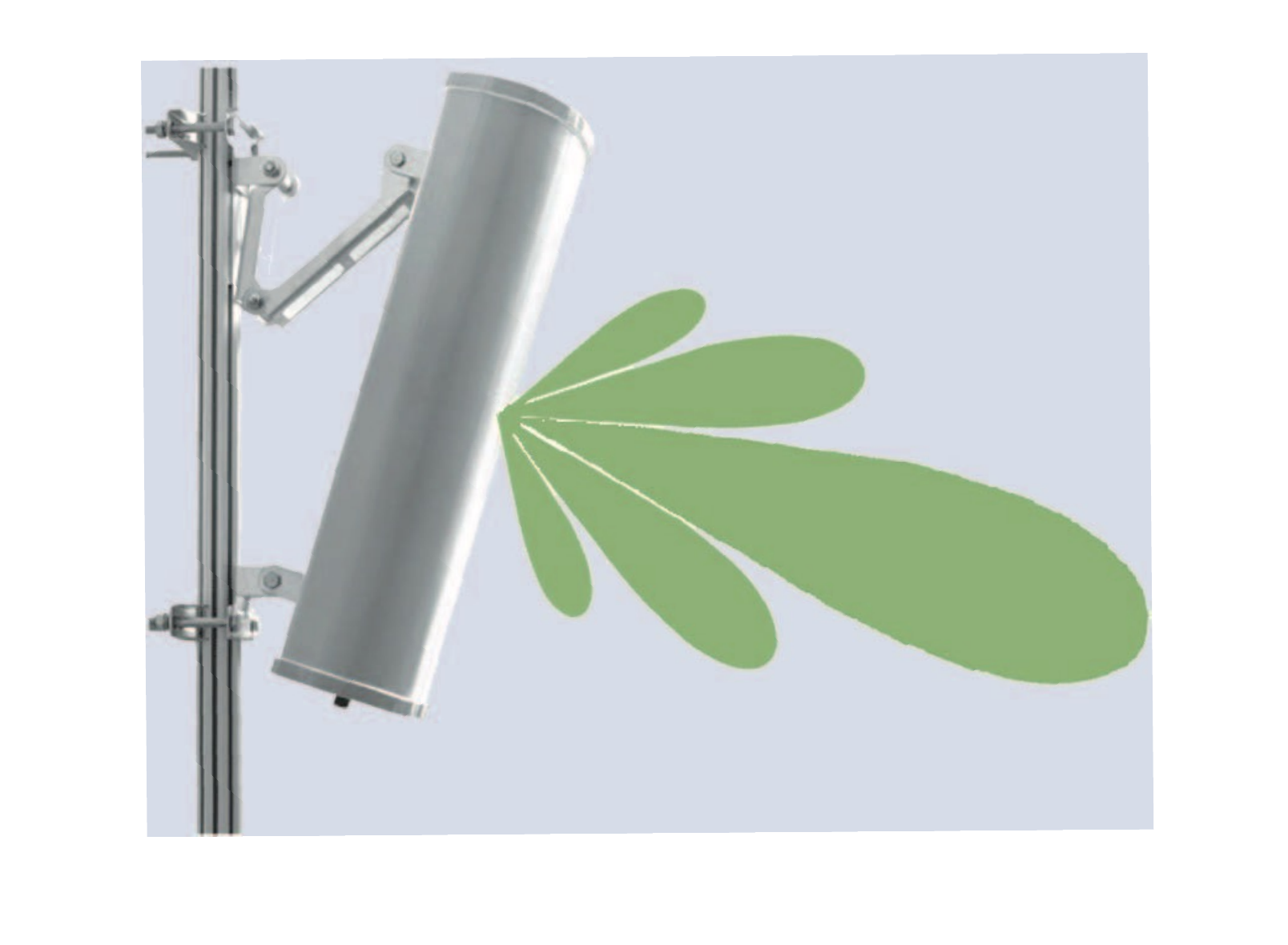}
		\end{minipage}}
		
	\subfigure[WiFi AP mounted on the ceil with $\theta_{tilt}=\frac{\pi}{2}$.]{
\begin{minipage}[t]{0.5\textwidth}
		\includegraphics[width=8cm]{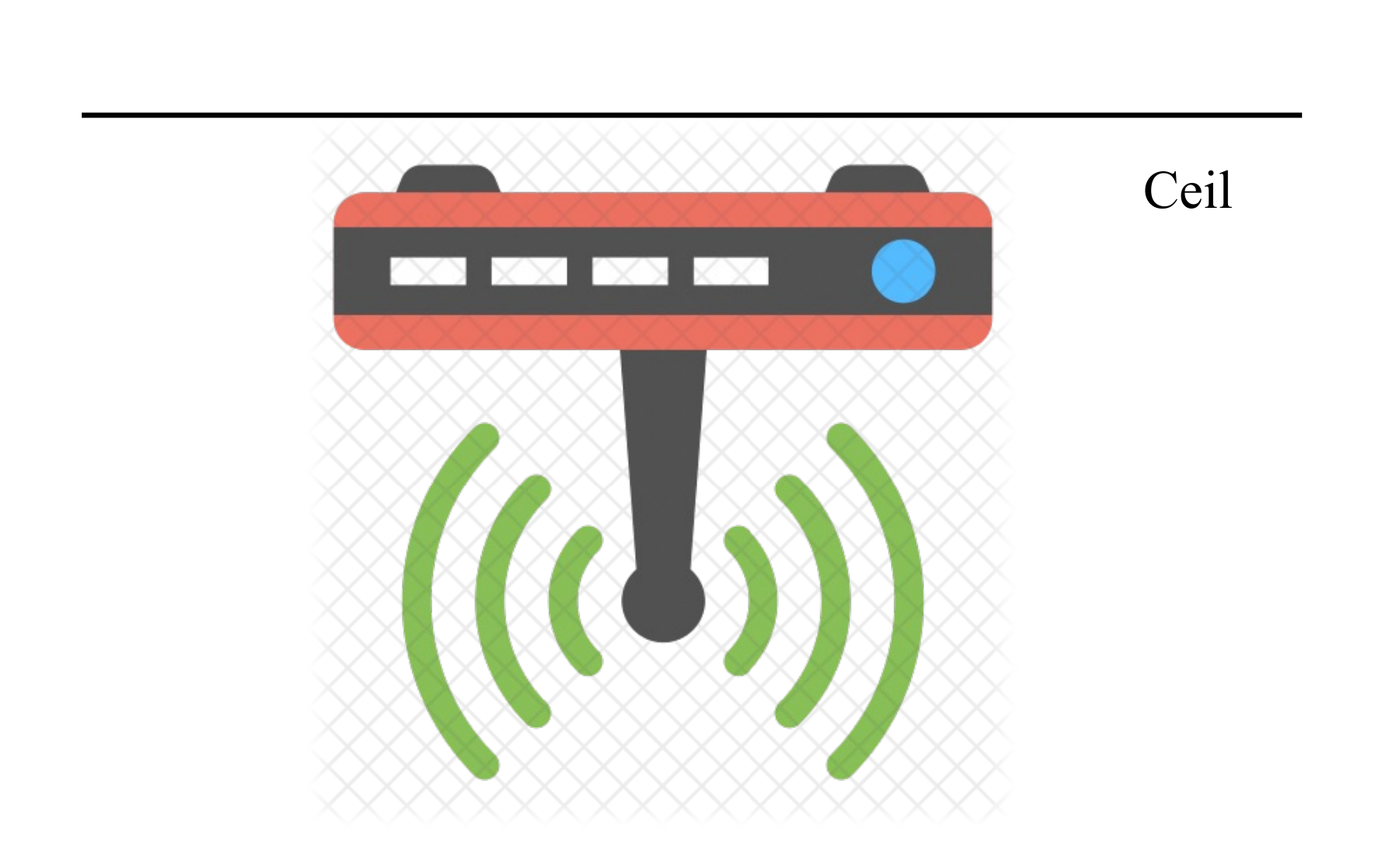}
		\end{minipage}}
\caption{Examples of integrated IRS-BS/AP.}\label{example}
\vspace{-10pt}\end{figure}

As shown in Fig. \ref{proposed_architecture}, we consider an integrated IRS-BS/AP architecture, where an antenna array and $J=4$ IRSs are deployed within a cuboid antenna radome, with $\mathcal J\triangleq\{1,2,3,4\}$ denoting the set of all IRSs. Specifically, the BS's antenna array is deployed at the center of the back surface of the antenna radome, while IRSs 1--4 are respectively deployed at its left, right, top, and bottom surfaces perpendicular to the antenna array. We consider a three-dimensional (3D) local  coordinate system (LCS) here, where the center of the antenna array is set as the origin $O$, and the antenna array is located on the $x$-$O$-$z$ plane. The antenna array is assumed to be a uniform planar array (UPA) with the size of $M=M_{x} \times M_{z}$, where $M_{x}$ and $M_{z}$ are the number of antennas along axes $x$ and $z$, respectively. The size of each IRS $j$ is denoted by $N_{j}=N_{j,1} \times N_{j,2}$, $j\in\mathcal J$, with $N_{j,1}$ and $N_{j,2}$ representing the number of reflecting elements along axes $y$ and $z$ (or $x$), respectively\footnote{In practice, since the antenna radome of cellular BSs or WiFi APs has a limited size (e.g., in the order of several to tens of carrier wavelengths), the number of IRS reflecting elements that can be integrated into the antenna radome is limited. However, the distance between the IRSs and the antenna array of the BS is small, which can significantly reduce the path loss of the reflected signals by IRSs as compared to conventional IRSs deployed much farther away from the BS/AP \cite{my,deploy,deploy2}. Moreover, we assume that the signal incident from the front half-space of each IRS is not blocked by the IRS on its opposite side, because in practice the size of the antenna radome along axis $y$ (i.e., thickness) is generally much smaller than that along axis $x$/$z$ (i.e., length/width) of the antenna radome, as shown in Fig. \ref{system_model}.}. The sets of elements for the BS UPA and IRS $j$, $j \in \mathcal{J}$, are denoted by $\mathcal M\triangleq \{1,2,\cdots,M\}$ and $\mathcal N_{j}\triangleq\{1,2,\cdots,N_{j}\}$, respectively. In the considered LCS, the location of the $m$-th antenna is denoted as $\mv s_{m}=[s_{m,x},s_{m,y},s_{m,z}]^{T}\in\mathbb{R}^{3\times 1},~ m\in\mathcal M$, and that of the $n_{j}$-th reflecting element of IRS $j$ is denoted as $\mv w_{j,n_{j}}=[w_{j,n_{j},x},w_{j,n_{j},y},w_{j,n_{j},z}]^{T}\in\mathbb{R}^{3\times 1},~j\in\mathcal J,~n_{j}\in\mathcal N_{j}$. Since the IRS can only reflect signals to/from its front half-space, each IRS should be deployed to face the antenna array and all other IRSs so as to efficiently reflect signals to/from them. Denote $\mv\delta_{j}$ as the normal vector of the plane which IRS $j$ is located in, with $\|\mv\delta_{j}\|=1,~j\in\mathcal J$. According to the orientation of the IRSs, we have $\mv\delta_{1}=[-1,0,0]^{T}$, $\mv\delta_{2}=[1,0,0]^{T}$, $\mv\delta_{3}=[0,0,-1]^{T}$, and $\mv\delta_{4}=[0,0,1]^{T}$. Moreover, as shown in Fig. \ref{system_model}, the antenna radome is deployed at altitude $H_{AR}$ with a suspension angle of $\theta_{tilt}\in[0,\pi/2]$ with respect to (w.r.t.) the vertical direction (which is perpendicular to the ground). Note that the suspension angle $\theta_{tilt}$ can be adjusted according to the requirement of specific communication applications, such as $\theta_{tilt}\in [0,\pi/12]$ for cellular BSs \cite{sector_bs} (see Fig. \ref{example}(a)) and $\theta_{tilt}=\pi/2$ for WiFi APs mounted on the ceiling (see Fig. \ref{example} (b)). In this paper, we focus on the uplink transmission from $K$ users to the BS, where the set of users is denoted by $\mathcal K\triangleq\{1,2,\cdots,K\}$.


\subsection{Channel Model}
Let $\vartheta_{j,n_j}$ denote the unit-modulus reflection coefficient for the $n_{j}$-th element of IRS $j$ with $|\vartheta_{j,n_{j}}|=1,~j\in\mathcal J, n_{j}\in\mathcal N_{j}$, and $\mv\Theta\triangleq\{\vartheta_{j,n_j},~j\in\mathcal J,n_j\in\mathcal N_{j}\}$. As shown in Fig. \ref{system_model}, the effective channel between each user and the BS's antenna array (i.e., UPA) in our proposed integrated IRS-BS architecture is a superposition of the direct channel component from the user to the antenna array of the BS without any IRS's reflection and the reflection channel components via IRSs. Due to the high path loss induced by multiple reflections among the IRSs, the power of the reflection channels is dominated by the single-reflection and double-reflection channel components. Thus, the effective channel from user $k$ to the antenna array of the BS is given by
\begin{align}\label{effective_channel}
	\mv h_{k}(\mv\Theta)=&\tilde{\mv h}_{k}+\underbrace{\sum_{j=1}^{J}\sum_{n_{j}=1}^{N_{j}}\mv f_{k,j}^{n_j}\vartheta_{j,n_{j}}}_{\text{single-reflection}} \nonumber\\
	&+\underbrace{\sum_{j=1}^{J}\sum_{q=1,q\neq j}^{J}\sum_{n_j=1}^{N_{j}}\sum_{n_q=1}^{N_q}\mv g_{k,j,q}^{n_{j},n_{q}}\vartheta_{j,n_{j}}\vartheta_{q,n_{q}}}_{\text{double-reflection}},
\end{align}
where $\tilde{\mv h}_{k}=[\tilde{h}_{k,1},\cdots,\tilde{h}_{k,M}]^{T}\in\mathbb{C}^{M\times 1}$ represents the direct channel component from user $k$ to the antenna array of the BS; $\mv f_{k,j,m}^{n_j} =[f_{k,j,1}^{n_j},\cdots,f_{k,j,M}^{n_j}]^{T}\in\mathbb{C}^{M\times 1}$ denotes the channel component from user $k$ to the antenna array of the BS via a single reflection by the $n_{j}$-th element of IRS $j$; and $\mv g_{k,j,q}^{n_{j},n_{q}} =[g_{k,j,q,1}^{n_{j},n_{q}},\cdots,g_{k,j,q,M}^{n_{j},n_{q}}]^{T}\in\mathbb{C}^{M\times 1}$ is the channel component from user $k$ to the antenna array of the BS via double reflections by the $n_{j}$-th element of IRS $j$ and subsequently the $n_{q}$-th element of IRS $q$. Next, we present the models of the direct, single-reflection, and double-reflection channel components, respectively.

\subsubsection{Direct Channel Component}
\begin{figure}
\centering
\includegraphics[width=8cm]{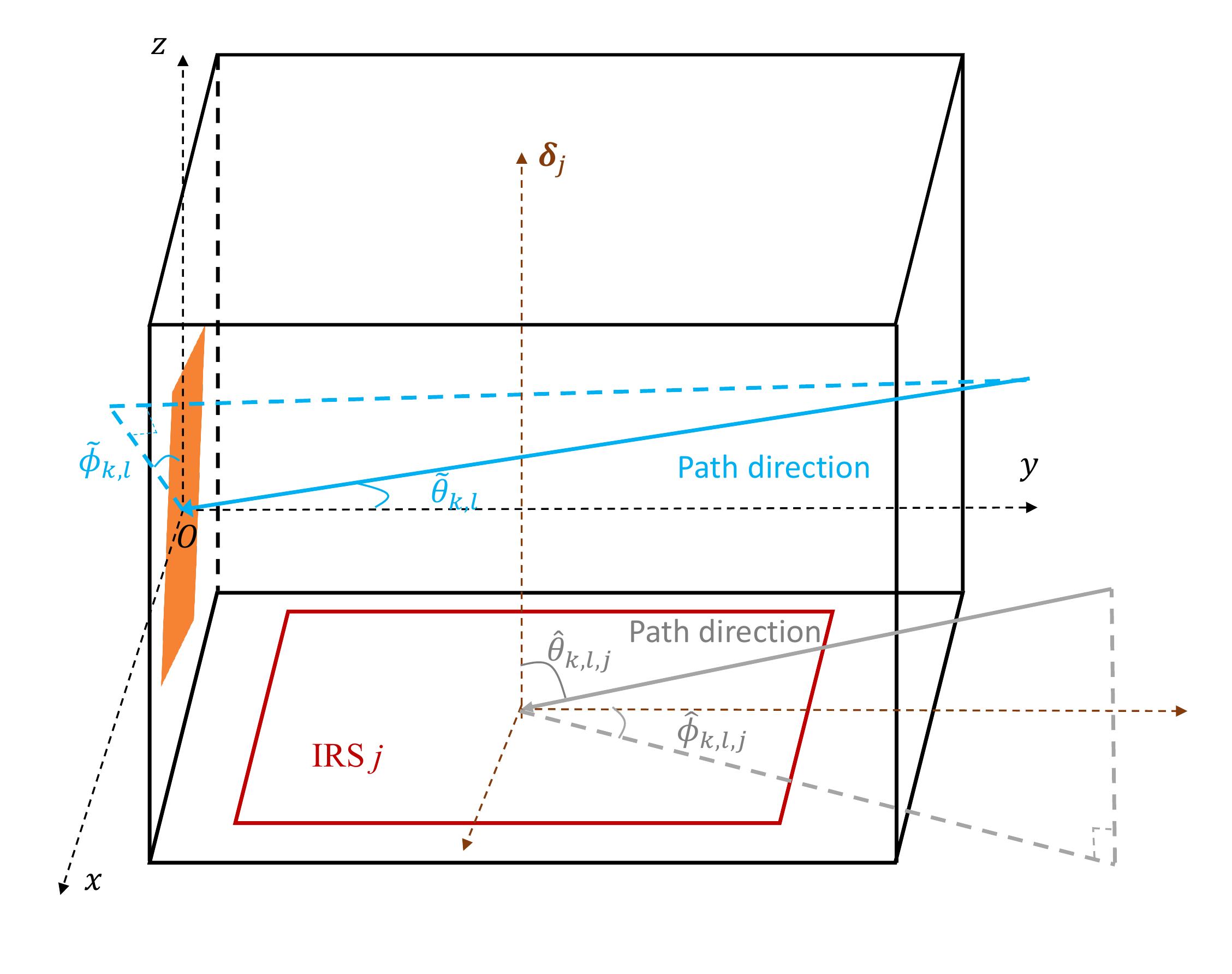}
\vspace{-10pt}\caption{Illustration of the elevation and azimuth angles w.r.t. the BS's antenna array and IRSs.}\label{angle_example}
\vspace{-10pt}\end{figure}
For the direct channel component from each user to the antenna array of the BS, the far-field condition is practically satisfied due to the much longer signal propagation distance between the user and BS w.r.t. the size of the BS's UPA. Thus, the UPW model can be used to characterize the channel response, where the one-dimensional (1D) array response vector can be defined as \cite{moving}
\begin{align}
	\mv e(\phi,\bar{M})=[1,e^{i\pi\phi},\cdots,e^{i(\bar{M}-1)\pi\phi}]^{T},\label{steering}
\end{align}
with $\phi$ denoting the phase difference (normalized to $\pi$) of the receive/transmit signals between two adjacent antenna elements, and $\bar{M}$ denoting the number of elements of interest in the 1D array. Thus, the direct channel component between user $k$ and the antenna array of the BS is given by
\begin{align}
	&\tilde{\mv h}_{k}=\sum_{l=1}^{L_{k}}\tilde{a}_{k,l}\sqrt{\tilde{G}(\tilde{\theta}_{k,l},\tilde{\phi}_{k,l})}
	\mv e\left(\frac{2d_{A}}{\lambda}\sin\tilde{\theta}_{k,l}\cos\tilde{\phi}_{k,l},M_{z}\right)\nonumber\\
	&\otimes \mv e\left(\frac{2d_{A}}{\lambda}\sin\tilde{\theta}_{k,l}\sin\tilde{\phi}_{k,l},M_{x}\right),\label{user_bs}
\end{align}
where $L_{k}$ is the number of (significant) signal propagation paths for user $k$,\footnote{In this paper, we consider the narrow-band channel model, while the channel dispersion effect is assumed to be negligible by applying multi-carrier modulation techniques, such as OFDM.} with $\mathcal {L}_{k}\triangleq\{1,2,\cdots, L_{k}\}$ denoting the set of all paths; $\tilde{a}_{k,l}\in\mathbb{C}$ denotes the complex gain of the $l$-th path from user $k$ to the BS; $\tilde{\theta}_{k,l}\in[0,\pi/2]$ and $\tilde{\phi}_{k,l}\in[0,2\pi)$ represent the elevation and azimuth AoAs of the $l$-th path w.r.t. the antenna array of the BS as shown in Fig. \ref{angle_example}; $d_{A}$ is the spacing between two adjacent antenna elements along axes $x$ and $z$; $\lambda$ denotes the carrier wavelength; and $\tilde{G}(\tilde{\theta}_{k,l},\tilde{\phi}_{k,l})$ is the antenna gain corresponding to the $l$-th path of user $k$, which is determined by the elevation and azimuth AoAs, $\tilde{\theta}_{k,l}$ and $\tilde{\phi}_{k,l}$, in general for any given antenna radiation pattern at the BS\footnote{Various types of BS/AP antenna radiation patterns can be used here, such as the isotropic antenna pattern, cosine-function based pattern, 3GPP antenna radiation patterns \cite{antenna_model}, etc.}.

\subsubsection{Single-Reflection Channel Component}
The single-reflection channel component is comprised of the link from the user to each IRS and the reflection link from that IRS to the antenna array of the BS. In our proposed integrated IRS-BS architecture, the far-field condition between the users and IRSs is satisfied due to the long signal propagation distance between them, where the UPW-based channel model can be applied, similar to the direct channel from each user to the BS's UPA. However, the distance between the IRSs and the antenna array of the BS is much shorter due to the limited size of the antenna radome and can be even smaller than the size of the UPA/IRSs; as a result, the UPW assumption between each IRS and the BS UPA does not hold. In this regard, we adopt the uniform spherical wave (USW) to model the element-wise channel from each IRS's reflecting element to the BS UPA (or equivalently, the UPW model for the channel from each IRS's reflecting element to each element of the BS's UPA) \cite{near_field1,near_field2}. Notice that the single-reflection channel component from a user to the antenna array of the BS is determined by the channels between the user and all IRSs' reflecting elements, the reflection gain provided by all IRSs' reflecting elements, the channels between all IRSs' reflecting elements and the BS antennas, and the antenna gain at the BS. Next, we present the models of the above elements in detail.

Without loss of generality, we define the $n_{j}$-th element of IRS $j$ as the one located at row $n_{j,1}$ ($1 \leq n_{j,1} \leq N_{j,1}$) and column $n_{j,2}$ ($1 \leq n_{j,2} \leq N_{j,2}$), where we define $n_{j}$ in terms of $(n_{j,1},n_{j,2})$ as $n_{j}=n_{j,1}+n_{j,2}(N_{j,1}-1)$. According to the UPW model, we can obtain the array response vector of the $l$-th path between user $k$ and IRS $j$ as $\mv{\alpha}_{k,l,j} \in \mathbb{C}^{N_j \times 1}$, with the $n_{j}$-th entry given by
\begin{align}
	&\alpha_{k,l,j}^{n_j}(\hat{\theta}_{k,l,j},\hat{\phi}_{k,l,j})=e^{i\varphi_{j}}\times\nonumber\\
	&e^{i\pi \frac{2d_{I}}{\lambda}\left[(n_{j,1}-1)\sin\hat{\theta}_{k,l,j}\cos\hat{\phi}_{k,l,j}+(n_{j,2}-1)\sin\hat{\theta}_{k,l,j}\sin\hat{\phi}_{k,l,j}\right]},\label{antenna_response}
\end{align}
where $\varphi_{j}$ denotes the initial phase of the signals received at IRS $j$ relative to the origin $O$; $d_{I}$ is the spacing between any two adjacent IRS reflecting elements; and $\hat{\theta}_{k,l,j}\in[0,\pi)$ and $\hat{\phi}_{k,l,j} \in [0, 2\pi)$ are the elevation and azimuth AoAs of the $l$-th path for user $k$ w.r.t. IRS $j$ as shown in Fig. \ref{angle_example}. According to geometry, we obtain the relation between the AoAs w.r.t. the antenna array and those w.r.t. the IRSs as
\begin{align}
	&\sin\hat{\theta}_{k,l,j}\cos\hat{\phi}_{k,l,j}=\cos\tilde{\theta}_{k,l},,~j=1,2,\\
	&\sin\hat{\theta}_{k,l,j}\sin\hat{\phi}_{k,l,j}=\sin\tilde{\theta}_{k,l}\cos\tilde{\phi}_{k,l},~j=1,2,\\
	&\sin\hat{\theta}_{k,l,j}\cos\hat{\phi}_{k,l,j}=\cos\tilde{\theta}_{k,l},~j=3,4,\\
	&\sin\hat{\theta}_{k,l,j}\sin\hat{\phi}_{k,l,j}=\sin\tilde{\theta}_{k,l}\sin\tilde{\phi}_{k,l},~j=3,4,\\
	&\cos\hat{\theta}_{k,l,1}=-\sin\tilde{\theta}_{k,l}\sin\tilde{\phi}_{k,l},\\
	&\cos\hat{\theta}_{k,l,2}=\sin\tilde{\theta}_{k,l}\sin\tilde{\phi}_{k,l},\label{cos1}\\
	&\cos\hat{\theta}_{k,l,3}=-\sin\tilde{\theta}_{k,l}\cos\tilde{\phi}_{k,l},\\
	&\cos\hat{\theta}_{k,l,4}=\sin\tilde{\theta}_{k,l}\cos\tilde{\phi}_{k,l}\label{cos2}.
\end{align}

Note that the electromagnetic response for each reflecting element of IRS is not isotropic for different incident/reflected angles in general. Thus, we define the reflection gain of each IRS element as \cite{channel_model}
 \begin{align}\label{reflection_gain}
	&G(\hat{\theta}^{A},\hat{\theta}^{D})=\nonumber\\
	&\begin{cases}
		2\times\frac{A\cos\hat{\theta}^{A}}{\lambda^{2}/4\pi}\times\frac{A\cos\hat{\theta}^{D}}{\lambda^{2}/4\pi},~&\text{if}~\hat{\theta}^{A}\in[0,\frac{\pi}{2})~\text{and}~\hat{\theta}^{D}\in[0,\frac{\pi}{2}),\\
		0,~&\text{otherwise},
	\end{cases}
\end{align}
where $\hat{\theta}^{A}$ and $\hat{\theta}^{D}$ denote the elevation AoA of the incident signals and the elevation AoD of the reflected signals w.r.t. the IRS reflecting element, respectively, factor $2$ is due to the half-space reflection of the IRS \cite{near_field3}, and $A$ is the area of each reflecting element. As can be observed from \eqref{reflection_gain}, each IRS element can only provide reflection gain when the signals are incident from and reflected to its front half-space, i.e., $\hat{\theta}^{A}\in[0,\frac{\pi}{2})$ and $\hat{\theta}^{D}\in[0,\frac{\pi}{2})$; otherwise, the reflection gain is set as zero. 

Let $\xi_{j,m}^{n_{j}}$ denote the complex channel gain between the $n_{j}$-th reflecting element of IRS $j$ and the $m$-th antenna at the BS. Since the sizes of each IRS reflecting element and antenna element are of sub-wavelength, the UPW assumption still holds for the LoS channel between any two IRS/antenna elements, despite their short distance \cite{antenna_book}. As such, we have
\begin{align}
	\xi_{j,m}^{n_j}=\frac{\lambda}{4\pi||\mv s_{m}-\mv w_{j,n_{j}}||}e^{-i\frac{2\pi}{\lambda}||\mv s_{m}-\mv w_{j,n_{j}}||}\label{irs_bs}.
\end{align}

Next, the single-reflection channel component from user $k$ to the $m$-th antenna of the BS via the $n_{j}$-th reflecting element of IRS $j$, $f_{k,j,m}^{n_j},~k\in\mathcal K,j\in\mathcal J, n_{j}\in\mathcal N_{j},m\in\mathcal M$, can be expressed as the summation of the single-reflection responses for the $L_k$ paths from user $k$, where the single-reflection response of each path is equal to the product of the channel gain between user $k$ and the $n_{j}$-th reflecting element of IRS $j$, the reflection gain provided by the $n_{j}$-th reflecting element of IRS $j$, the channel gain between the $n_{j}$-th reflecting element of IRS $j$ and the $m$-th antenna, and the antenna gain of the $m$-th antenna, i.e., 
\begin{align}\label{single_reflection}
	&f_{k,j,m}^{n_j}=\sum_{l=1}^{L_{k}}\tilde{a}_{k,l}\alpha_{k,l,j}^{n_j}(\hat{\theta}_{k,l,j},\hat{\phi}_{k,l,j}) \sqrt{G(\hat{\theta}_{k,l,j},\hat{\theta}_{j,m}^{n_j})}\nonumber\\
	&\times \xi_{j,m}^{n_j}\sqrt{\tilde{G}(\tilde{\theta}_{m,j}^{n_{j}},\tilde{\phi}_{m,j}^{n_j})},
\end{align}
where $\tilde{a}_{k,l}\in\mathbb{C}$ is the complex gain of the $l$-th path from user $k$ to the BS given in \eqref{user_bs}; $G(\hat{\theta}_{k,l,j},\hat{\theta}_{j,m}^{n_j})$ denotes the reflection gain provided by the $n_{j}$-th reflecting element of IRS $j$ as defined in (\ref{reflection_gain}), with $\hat{\theta}_{j,m}^{n_j}=\arccos\frac{(\mv s_m-\mv w_{j,n_j})^{T}\mv\delta_{j}}{||\mv s_{m}-\mv w_{j,n_j}||}$ being the elevation angle of the direction of the $m$-th antenna w.r.t. the $n_j$-th reflecting element of IRS $j$; and $\tilde{G}(\tilde{\theta}_{m,j}^{n_j},\tilde{\phi}_{m,j}^{n_j})$ denotes the antenna gain provided by the $m$-th antenna as defined in \eqref{user_bs}, with $\tilde{\theta}_{m,j}^{n_j}=\arccos \frac{w_{j,n_j,y}-s_{m,y}}{||\mv w_{j,n_j}-\mv s_{m}||}$ and $\tilde{\phi}_{m,j}^{n_j}=\arctan \frac{w_{j,n_j,x}-s_{m,x}}{w_{j,n_j,z}-s_{m,z}}$ being the elevation and azimuth angles of the direction of the $n_j$-th reflecting element of IRS $j$ w.r.t. the $m$-th antenna, respectively.

\subsubsection{Double-Reflection Channel Component}
Similar to the single-reflection channel component, the double-reflection channel component from user $k$ to the $m$-th antenna of the BS via the $n_{j}$-th reflecting element of IRS $j$ and then the $n_q$-th reflecting element of IRS $q$, $g_{k,j,q,m}^{n_{j},n_{q}},~k\in\mathcal K, j\neq q\in\mathcal J, n_{j}\in\mathcal N_{j},n_{q}\in\mathcal N_{q},m\in\mathcal M$, is the summation of the double-reflection responses for the $L_k$ paths from user $k$. The double-reflection response of each path is given by the product of the channel gain between user $k$ and the $n_{j}$-th reflecting element of IRS $j$, the reflection gain provided by the $n_{j}$-th reflecting element of IRS $j$, the channel gain between the $n_{j}$-th reflecting element of IRS $j$ and the $n_{q}$-th reflecting element of IRS $q$, the reflection gain provided by the $n_{q}$-th reflecting element of IRS $q$, the channel gain between the $n_{q}$-th reflecting element of IRS $q$ and the $m$-th antenna, and the antenna gain of the $m$-th antenna, i.e., 
\begin{align}\label{double_reflection}
&g_{k,j,q,m}^{n_j,n_q}=\sum_{l=1}^{L_{k}}\tilde{a}_{k,l}\alpha_{k,l,j}^{n_j}(\hat{\theta}_{k,l,j},\hat{\phi}_{k,l,j})\sqrt{G(\hat{\theta}_{k,l,j},\hat{\theta}_{j,q}^{n_j,n_q})} \nonumber \\
&~~~~~~~~\times \zeta_{j,q}^{n_j,n_q} \sqrt{G(\hat{\theta}_{q,j}^{n_q,n_j},\hat{\theta}_{q,m}^{n_q})} \xi_{q,m}^{n_q} \sqrt{\tilde{G}(\tilde{\theta}_{m,q}^{n_q},\tilde{\phi}_{m,q}^{n_q})},
\end{align}
where
\begin{align}
\zeta_{j,q}^{n_j,n_q}=\frac{\lambda}{4\pi||\mv w_{q,n_{q}}-\mv w_{j,n_{j}}||}e^{-i\frac{2\pi}{\lambda}||\mv w_{q,n_{q}}-\mv w_{j,n_{j}}||}\label{inter_irs}
\end{align}
denotes the complex channel gain for the LoS path between the $n_j$-th reflecting element of IRS $j$ and the $n_q$-th reflecting element of IRS $q$ due to their short distance; $G(\hat{\theta}_{k,l,j},\hat{\theta}_{j,q}^{n_j,n_q})$ is the reflection gain provided by the $n_j$-th reflecting element of IRS $j$, with 
$\hat{\theta}_{j,q}^{n_j,n_q}=\arccos \frac{(\mv w_{q,n_q}-\mv w_{j,n_j})^{T}\mv\delta_{j}}{||\mv w_{q,n_q}-\mv w_{j,n_j}||}$ denoting the elevation angle of the direction of the $n_q$-th reflecting element of IRS $q$ w.r.t. the $n_j$-th reflecting element of IRS $j$; $G(\hat{\theta}_{q,j}^{n_q,n_j},\hat{\theta}_{q,m}^{n_q})$ is the reflection gain provided by the $n_q$-th reflecting element of IRS $q$, with $\hat{\theta}_{q,j}^{n_q,n_j}=\arccos \frac{(\mv w_{j,n_j}-\mv w_{q,n_q})^{T}\mv\delta_{q}}{||\mv w_{j,n_j}-\mv w_{q,n_q}||}$ and $\hat{\theta}_{q,m}^{n_q}=\arccos \frac{(\mv s_m-\mv w_{q,n_q})^{T}\mv\delta_{q}}{||\mv s_{m}-\mv w_{q,n_q}||}$ denoting the elevation angles of the directions of the $n_j$-th reflecting element of IRS $j$ and the $m$-th antenna w.r.t. the $n_q$-th reflecting element of IRS $q$; and $\tilde{G}(\tilde{\theta}_{m,q}^{n_q},\tilde{\phi}_{m,q}^{n_q})$ is the antenna gain provided by $m$-th antenna, with $\tilde{\theta}_{m,q}^{n_q}=\arccos\frac{w_{q,n_q,y}-s_{m,y}}{||\mv w_{q,n_q}-\mv s_{m}||}$ and $\tilde{\phi}_{m,q}^{n_q}=\arctan \frac{w_{j,n_j,x}-s_{m,x}}{w_{j,n_j,z}-s_{m,z}}$ denoting the elevation and azimuth angles of the direction of the $n_q$-th reflecting element of IRS $q$ w.r.t. the $m$-th antenna, respectively. For ease of reading, the main symbol notations used in this paper and their physical meanings are summarized in Table \ref{notations}.

\begin{spacing}{1.0}
\begin{table*}
\centering
\caption{Symbols and Physical Meanings}\label{notations}
\small
\begin{tabular}{|p{1.5cm}|p{5.5cm}||p{2.5cm}|p{5.5cm}|}
\hline
Symbol&Physical meaning&Symbol&Physical meaning\\
\hline
\hline
$M$&Number of BS antennas&$J$&Number of IRSs\\
\hline
$N_{j}$&Number of reflecting elements of IRS $j$&$N$&Total number of reflecting elements of all IRSs\\
\hline
$M_{x}$/$M_{z}$&Number of antennas along axis $x$ or axis $z$  &$N_{j,1}$/$N_{j,2}$&Number of reflecting elements along axis $y$ or axis $z$($x$) for IRS $j$\\
\hline
$K$&Number of users&$\mv s_{m}$&Coordinates of the $m$-th antenna\\
\hline
$\mv w_{j,n_{j}}$&Coordinates of the $n_{j}$-th reflecting element of IRS $j$&$\mv\delta_{j}$&Normal vector of IRS $j$\\
\hline
$\vartheta_{j,n_j}$&Reflection coefficient of the $n_j$-th reflecting element of IRS $j$&$H_{AR}$&Altitude of the antenna radome\\
\hline
$\theta_{tilt}$&Suspension angle of the BS antenna radome&$A$&Area of each reflecting element\\
\hline
$\lambda$&Carrier wavelength&$d_{A}$/$d_{I}$&Spacing between two adjacent elements of the antenna array/IRSs\\
\hline
$\tilde{\theta}_{m,j}^{n_j}$/$\tilde{\phi}_{m,j}^{n_j}$&Elevation/azimuth angle of the direction of the $n_j$-th reflecting element of IRS $j$ w.r.t. the $m$-th BS antenna&$\hat{\theta}_{j,m}^{n_j}$/$\hat{\theta}_{j,q}^{n_j,n_q}$&Elevation angle of the direction of the $m$-th antenna/the $n_q$-th reflecting element of IRS $q$ w.r.t. the $n_j$-th reflecting element of IRS $j$\\
\hline
$\tilde{G}(\cdot)$&Antenna gain of the BS antennas&$G(\cdot)$&Reflection gain of the IRS element\\
\hline
$\mv f_{k,j}^{n_{j}}$&Channel component from user $k$ to the antenna array of the BS via a single reflection at the $n_j$-th element of IRS $j$&$\mv g_{k,j,q}^{n_{j},n_{q}}$&Channel component from user $k$ to the antenna array of the BS via double reflections at the $n_j$-th element of IRS $j$ and then the $n_{q}$-th reflecting element of IRS $q$\\
\hline
$\xi_{k,j,m}^{n_j}$&Complex channel gain between the $n_j$-th reflecting element of IRS $j$ and the $m$-th BS antenna&$\zeta_{j,q}^{n_j,n_q}$&Complex channel gain between the $n_j$-th reflecting element of IRS $j$ and the $n_q$-th reflecting element of IRS $q$\\
\hline
$\varphi_{j}$&Initial phase of the received signals at IRS $j$ relative to the origin $O$&$L_k$&Number of paths from user $k$\\
\hline
$\tilde{a}_{k,l}$&Complex path gain of the $l$-th path from user $k$&$\alpha_{k,l,j}^{n_j}(\hat{\theta}_{k,l,j},\hat{\phi}_{k,l,j})$&Array response of the $n_{j}$-th reflecting element of IRS $j$ for the $l$-th path from user $k$\\
\hline
$\tilde{\theta}_{k,l}$/$\tilde{\phi}_{k,l}$&Elevation/azimuth AoA of the $l$-th path from user $k$ w.r.t. the BS antenna array&$\hat{\theta}_{k,l,j}$/$\hat{\phi}_{k,l,j}$&Elevation/azimuth AoA of $l$-th path from user $k$ w.r.t. IRS $j$\\
 \hline
 $P$&Transmit power of each user&$\sigma^{2}$&Noise power per BS antenna\\
 \hline
 $\eta$&Number of antenna modules in generalized IRS-BS architecture with modular antenna array&&\\
 \hline
\end{tabular}
\end{table*}
\end{spacing}

\subsection{Generalized IRS-BS Architecture With Modular Antenna Arrays}
\begin{figure}
\centering
\includegraphics[width=8cm]{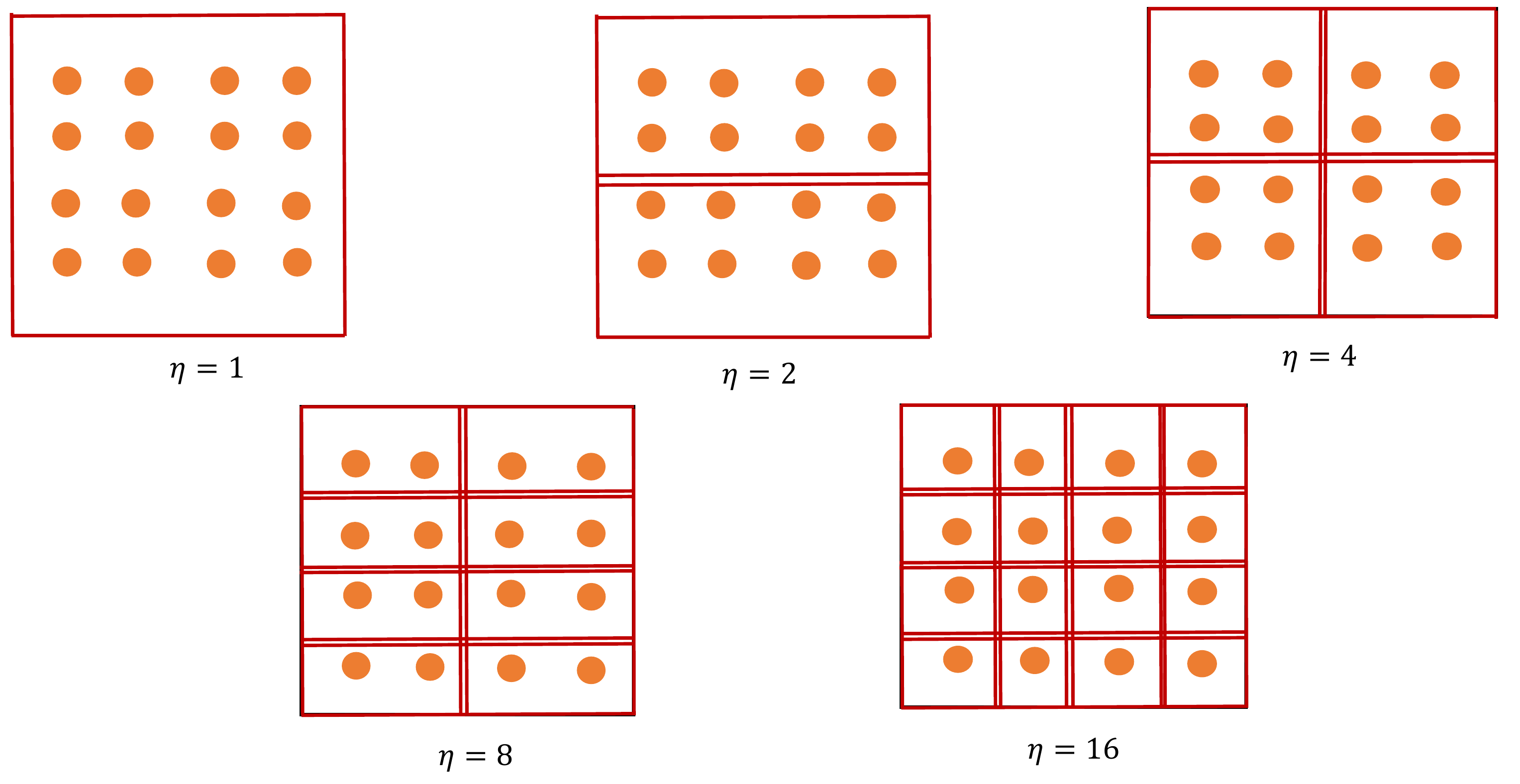}
\vspace{-10pt}\caption{Illustration of the generalized integrated IRS-BS architecture with modular antenna arrays, where the yellow circles represent antennas and the red lines indicate the surfaces to deploy IRS reflecting elements.}\label{modular_antenna_array}
\vspace{-20pt}\end{figure}

In this subsection, we generalize the proposed integrated IRS-BS architecture by employing modular antenna arrays, so that more reflecting elements can be accommodated inside the antenna radome. Specifically, we divide the antenna array into $\eta\geq 1$ modules, where each module consisting of $M/\eta$ antennas (assumed to be an integer) is surrounded by four smaller-size IRSs parallel to the left, right, top, and bottom surfaces. Note that the integrated IRS-BS architecture proposed in Section \ref{system_architecture}-A can be considered as a special case of the generalized IRS-BS architecture by setting $\eta=1$. However, the integrated IRS-BS architecture proposed in Section \ref{system_architecture}-A does not need to change the structure of existing antenna arrays and thus has lower implementation cost compared to the generalized IRS-BS architecture with $\eta>1$. Fig. \ref{modular_antenna_array} shows an example of the generalized IRS-BS architecture with modular antenna arrays, where the number of BS antennas is set as $M=16$ with $M_{x}=M_{z}=4$. It can be seen that the total number of reflecting elements increases as $\eta$ increases, e.g., $2N$ elements for $\eta=4$ and $4N$ elements for $\eta=16$. However, the number of reflecting elements surrounding each antenna module decreases as $\eta$ increases, e.g., $N/2$ elements for $\eta=4$ and $N/4$ for $\eta=16$. Note that the channels between the users and each antenna module for $\eta>1$ can be modelled in a similar way to that for $\eta=1$ as shown in the previous subsection, while we assume that there are no signal reflections among the reflecting elements surrounding different antenna modules\footnote{On one hand, since each IRS can only receive/reflect the signals from/towards its front half-space, signal reflections do not exist between two IRSs that are not located in the front half-space of each other. On the other hand, for two IRSs located at different modules, even though they are in the front half-space of each other, the signal reflections between them are negligible because they are blocked by other IRSs between them.}. In particular, as the number of antenna modules (i.e., $\eta$) increases, the number of single-reflection and double-reflection channel components between any user and each antenna element decreases, while the path loss of such reflection channel components is also reduced due to shorter signal propagation distance among the reflecting elements surrounding the same antenna module. The performance comparison between the generalized IRS-BS architectures under different numbers of antenna modules will be provided in Section \ref{numerical_results} via simulations.

\section{Problem Formulation and Passive Reflection Design}\label{reflection_design}
In this section, we formulate and solve the problem for optimizing the reflection coefficients of all IRSs in the integrated IRS-BS architecture presented in Section \ref{system_architecture}-A, while the results are also applicable to the generalized IRS-BS architecture with modular antenna arrays given in Section \ref{system_architecture}-C. 

\subsection{Problem Formulation}
Let $P$ denote the maximum transmit power of each user. By employing the minimum mean square error (MMSE) combining and successive interference cancellation (SIC) techniques at the BS, the sum-rate for decoding the signals from all the $K$ users in bits per second per Hertz (bps/Hz) is given by \cite{tse}
\begin{align}
R(\mv\Theta)=\log_{2}\text{det}\left(\mv I_{M}+\sum_{k=1}^{K}\frac{P}{\sigma^{2}}\mv h_{k}(\mv\Theta)\mv h_{k}(\mv\Theta)^{H}\right),\label{sum_rate}
\end{align}
where $\sigma^{2}$ denotes the independent CSCG noise power at each of the receive antennas at the BS.

In this paper, we aim to maximize the sum-rate of the $K$ users by optimizing the passive reflection coefficients of all the IRSs, i.e., $\mv\Theta$. The corresponding problem is formulated as
\begin{align}
\text{(P1):}~\max_{\mv\Theta}~&R(\mv\Theta)\nonumber\\
\text{s.t.}~~&|\vartheta_{j,n_{j}}|=1,~\forall j\in\mathcal J,n_{j}\in\mathcal N_{j}.\label{unit}
\end{align}
Problem (P1) is difficult to solve for the following reasons. On one hand, the CSI acquisition is a non-trivial task in practice because of much unknown channel parameters in $\{\tilde{\mv h}_{k}\}$, $\{\mv f_{k,j}^{n_{j}}\}$, and $\{\mv g_{k,j,q}^{n_{j},n_{q}}\}$, which may result in high overhead for channel estimation. On the other hand, problem (P1) is a non-convex optimization problem due to the non-concave objective function and unit-modulus constraints in (\ref{unit}). Besides, the passive reflection coefficients of all reflecting elements at all IRSs are coupled intricately in the objective function due to both the single- and double-reflection signals.  In the following, we consider two cases with/without perfect CSI at the BS to solve (P1), respectively. 

\subsection{Passive Reflection Design With Perfect CSI}
In this subsection, we assume that the perfect CSI for $\{\tilde{\mv h}_{k}\}$, $\{\mv f_{k,j}^{n_{j}}\}$, and $\{\mv g_{k,j,q}^{n_{j},n_q}\}$ is available at the BS by exploiting the channel estimation methods proposed in IRS-aided communications \cite{channel_estimation_survey,group,liuliang,anchor,sparsity,double_estimation}, where the obtained solution provides an upper bound on the achievable sum-rate of the users. To tackle the non-convex problem (P1), in the following, we propose an efficient algorithm by exploiting  the successive refinement method \cite{array,double2}, where we optimize each of the $N$ reflection coefficients with the other $(N-1)$ reflection coefficients being fixed in an iterative manner over the $N$ reflection coefficients.

For the given $\vartheta_{q,n_q},~q\neq j\in\mathcal J,n_{q}\in\mathcal N_{q}$, and $\vartheta_{j,\hat{n}_{j}},~\hat{n}_{j}\neq n_{j}\in\mathcal N_{j}$, the objective function of (P1) w.r.t. $\vartheta_{j,n_j}$ can be re-written as
\begin{align}
R_{j,n_j}(\vartheta_{j,n_j})&\triangleq\log_{2}\text{det}\Big(\mv A_{j,n_j}+\vartheta_{j,n_j}\mv B_{j,n_j}\nonumber\\
&+\vartheta_{j,n_j}^{*}\mv B_{j,n_j}^{H}\Big),~j\in\mathcal J, n_{j}\in\mathcal N_{j},
\end{align}
where $\mv A_{j,n_j}$ and $\mv B_{j,n_j}$ are constants independent to $\vartheta_{j,n_j}$ and respectively given by

\vspace{-10pt}\begin{small}\begin{align}
&\mv A_{j,n_j}=\mv I_{M}\nonumber\\
&+\frac{P}{\sigma^{2}}\sum_{k=1}^{K}\Big(\Big(\mv h_{k}+\sum_{\hat{n}_{j}\neq n_j}^{N_{j}}\mv f_{k,j}^{\hat{n}_{j}}\vartheta_{j,\hat{n}_{j}}+\sum_{q\neq j}\sum_{n_{q}=1}^{N_{q}}\mv f_{k,q}^{n_{q}}\vartheta_{q,n_{q}}\nonumber\\
&+\sum_{\hat{n}_{j}\neq n_{j}}^{N_{j}}\sum_{q\neq j}\sum_{n_{q}=1}^{N_{q}}\Big(\mv g_{k,j,q}^{\hat{n}_{j},n_{q}}+\mv g_{k,q,j}^{n_{q},\hat{n}_{j}}\Big)\vartheta_{j,\hat{n}_{j}}\vartheta_{q,n_{q}}\nonumber\\
&+\sum_{q\neq j}\sum_{r\neq q,r\neq j}\sum_{n_{q}=1}^{N_{q}}\sum_{n_{r}=1}^{N_{r}}\mv g_{k,q,r}^{n_{q},n_{r}}\vartheta_{q,n_{q}}\vartheta_{r,n_r}\Big)\nonumber\\
&\times\Big(\mv h_{k}+\sum_{\hat{n}_{j}\neq n_j}^{N_{j}}\mv f_{k,j}^{\hat{n}_{j}}\vartheta_{j,\hat{n}_{j}}+\sum_{q\neq j}\sum_{n_{q}=1}^{N_{q}}\mv f_{k,q}^{n_{q}}\vartheta_{q,n_{q}}\nonumber\\
&+\sum_{\hat{n}_{j}\neq n_{j}}^{N_{j}}\sum_{q\neq j}\sum_{n_{q}=1}^{N_{q}}\Big(\mv g_{k,j,q}^{\hat{n}_{j},n_{q}}+\mv g_{k,q,j}^{n_{q},\hat{n}_{j}}\Big)\vartheta_{j,\hat{n}_{j}}\vartheta_{q,n_{q}}\nonumber\\
&+\sum_{q\neq j}\sum_{r\neq q,r\neq j}\sum_{n_{q}=1}^{N_{q}}\sum_{n_{r}=1}^{N_{r}}\mv g_{k,q,r}^{n_{q},n_{r}}\vartheta_{q,n_{q}}\vartheta_{r,n_r}\Big)^{H}\nonumber\\
&+\Big(\mv f_{k,j}^{n_j}+\sum_{q\neq j}\sum_{n_q=1}^{N_q}\Big(\mv g_{k,j,q}^{n_j,n_q}+\mv g_{k,q,j}^{n_q,n_j}\Big)\vartheta_{q,n_q}\Big)\nonumber\\
&\times\Big(\mv f_{k,j}^{n_j}+\sum_{q\neq j}\sum_{n_q=1}^{N_q}(\mv g_{k,j,q}^{n_j,n_q}+\mv g_{k,q,j}^{n_q,n_j})\vartheta_{q,n_q}\Big)^{H}\Big),~j\in\mathcal J,n_{j}\in\mathcal N_{j},\label{A}\end{align}
\end{small}and

\vspace{-20pt}\begin{small}\begin{align}
&\mv B_{j,n_j}=\frac{P}{\sigma^{2}}\sum_{k=1}^{K}\Big(\Big(\mv f_{k,j}^{n_j}+\sum_{q\neq j}\sum_{n_q=1}^{N_q}\Big(\mv g_{k,j,q}^{n_j,n_q}+\mv g_{k,q,j}^{n_q,n_j}\Big)\vartheta_{q,n_q}\Big)\nonumber\\
&\times\Big(\mv h_{k}+\sum_{\hat{n}_{j}\neq n_j}^{N_{j}}\mv f_{k,j}^{\hat{n}_{j}}\vartheta_{j,\hat{n}_{j}}+\sum_{q\neq j}\sum_{n_{q}=1}^{N_{q}}\mv f_{k,q}^{n_{q}}\vartheta_{q,n_{q}}\nonumber\\
&+\sum_{\hat{n}_{j}\neq n_{j}}^{N_{j}}\sum_{q\neq j}\sum_{n_{q}=1}^{N_{q}}\Big(\mv g_{k,j,q}^{\hat{n}_{j},n_{q}}+\mv g_{k,q,j}^{n_{q},\hat{n}_{j}}\Big)\vartheta_{j,\hat{n}_{j}}\vartheta_{q,n_{q}}\nonumber\\
&+\sum_{q\neq j}\sum_{r\neq q,r\neq j}\sum_{n_{q}=1}^{N_{q}}\sum_{n_{r}=1}^{N_{r}}\mv g_{k,q,r}^{n_{q},n_{r}}\vartheta_{q,n_{q}}\vartheta_{r,n_r}\Big)^{H}\Big),~\forall j\in\mathcal J,n\in\mathcal N_{j}.\label{B}
\end{align}\end{small}Therefore, the subproblem for optimizing $\vartheta_{j,n_{j}}$ can be expressed as
\begin{align}
\text{(P1-$j$-$n_j$):}~\max_{\vartheta_{j,n_j}}~&R_{j,n_j}(\vartheta_{j,n_j})\nonumber\\
\text{s.t.}~&|\vartheta_{j,n_j}|= 1.\label{unit1}
\end{align}
Although the objective function of (P1-$j$-$n_j$) is concave w.r.t. $\vartheta_{j,n_j}$, it is still non-convex due to the unit-modulus constraint in (\ref{unit1}). To address this issue, we relax it as $|\vartheta_{j,n_j}|\leq 1$ and denote the relaxed problem as (P2-$j$-$n_j$), which is convex and thus can be solved by standard convex optimization methods, such as the interior point method \cite{convex}. Denote $\vartheta_{j,n_j}^{\star}$ as the optimal solution for (P2-$j$-$n_j$), which may not be feasible to (P1-$j$-$n_j$) due to the unit-modulus constraint in (\ref{unit1}). As such, the obtained solution should be projected to the feasible region of (P1-$j$-$n_j$) as follows \cite{projection}
\begin{align}
\vartheta^{\star\star}_{j,n_j}=\vartheta_{j,n_j}^{\star}/|\vartheta_{j,n_j}^{\star}|.\label{update}
\end{align}

With the solution for (P1-$j$-$n_j$) derived in (\ref{update}), we are ready to complete our proposed solution to solve (P1) with perfect CSI. Specifically, we first randomly generate $T\geq 1$ solutions of $\{\vartheta_{j,n_j},~j\in\mathcal J,n_j\in\mathcal N_{j}\}$ satisfying $|\vartheta_{j,n_j}|=1$, where the phase shifts of $\vartheta_{j,n_j}$'s follow the uniform distribution in $[0,2\pi)$. Then, we select the solution achieving the maximum sum-rate as the initial point. Subsequently, we update the reflection coefficients of all reflecting elements based on (\ref{update}) sequentially from $j=1$ and $n_{j}=1$ to $j=J$ and $n_{j}=N_{J}$. After the reflection coefficients of all reflecting elements are updated, we check whether the increment of the objective function in (P1) is smaller than a given small positive value $\epsilon_{1}$ or the maximum number of iterations $I_{\max}$ is reached. If this is the case, the algorithm terminates; otherwise, the update for the reflection coefficients of all reflecting elements is repeated as above. The overall solution to solve problem (P1) with perfect CSI is summarized in Algorithm \ref{successive}. Denote $I$ as the number of iterations required to update the reflection coefficients of all IRSs in Algorithm \ref{successive} (i.e., line 3--line 11) with $I\leq I_{\max}$. Notice that the reflection coefficient of each reflecting element is updated by applying the standard interior-point method to solve problem (P2-$j$-$n_j$), which incurs the complexity of $\mathcal{O}(3K^{3}+2K^{2}M)$ \cite{complex}, and the calculation for sum-rate has the complexity of $\mathcal{O}(M^{2})$. As a result, the overall computational complexity of Algorithm \ref{successive} is $\mathcal {O}(TM^{2}+IN(3K^{3}+2K^{2}M)+IM^{2})$.

\begin{algorithm}
\caption{Passive Reflection Design with Perfect CSI.}
\label{successive}
\begin{algorithmic}[1]
\REQUIRE {$\{\tilde{\mv h}_{k}\}$, $\{\mv f_{k,j}^{n_j}\}$, $\{\mv g_{k,j,q}^{n_j,n_q}\}$, $P$, $\sigma^{2}$, and $T$.}
\STATE {Randomly generate $T$ independent realizations of $\{\vartheta_{j,n_j},~j\in\mathcal J,n_j\in\mathcal N_{j}\}$, where the phase shift of each $\vartheta_{j,n_j}$ is randomly chosen from interval $[0,2\pi)$.}
\STATE{Select $\{\tilde{\vartheta}_{j,n_j}^{\star},~j\in\mathcal J,n_j\in\mathcal N_{j}\}$ as the realization yielding the largest objective value of (P1), and initialize $\vartheta_{j,n_j}=\tilde{\vartheta}_{j,n_j}^{\star},~j\in\mathcal J,n_j\in\mathcal N_{j}$. }
\REPEAT
\FOR{$j=1\rightarrow J$}
\FOR{$n_j=1\rightarrow N_j$}
\STATE{For the given $\vartheta_{q,n_q},~q\neq j\in\mathcal J,n_{q}\in\mathcal N_{q}$, and $\vartheta_{j,\hat{n}_{j}},~\hat{n}_{j}\neq n_{j}\in\mathcal N_{j}$, obtain $A_{j,n_j}$ and $B_{j,n_j}$ according to (\ref{A}) and (\ref{B}).}
\STATE{Obtain $\vartheta_{j,n_j}^{\star}$ via solving (P2-$j$-$n_{j}$) and update $\vartheta^{\star\star}_{j,n_{j}}$ according to (\ref{update}).}
\STATE{Set $\vartheta_{j,n_j}=\vartheta^{\star\star}_{j,n_j}$. }
\ENDFOR
\ENDFOR
\UNTIL{the increment of the objective function in (P1) is smaller than $\epsilon_{1}$ or the maximum number of iterations $I_{\max}$ is reached. }
\ENSURE {$\{\vartheta_{j,n_j},~j\in\mathcal J,n_j\in\mathcal N_{j}\}$.}
\end{algorithmic}
\end{algorithm}

\subsection{Passive Reflection Design Without CSI}
In practice, the CSI acquisition for the considered integrated IRS-BS architecture is a challenging problem because the direct channel components $\{\tilde{\mv h}_{k}\}$, single-reflection channel components $\{\mv f_{k,j}^{n_{j}}\}$, and double-reflection channel components $\{\mv g_{k,j,q}^{n_{j},n_q}\}$ involve $KM+KMN+KM\sum_{j=1}^{J}\sum_{q\neq j}N_{j}N_{q}$ unknown channel parameters in total, which may result in extremely high training overhead if the number of reflecting elements (i.e., $N$) is large, and thus a substantial degradation of the achievable rates of the users for data transmission. Moreover, the conventional transmission protocol should be reformulated for the additional channel estimation of reflection channel components. To address these issues, in this subsection, we propose practical methods for designing the passive reflection coefficients of all IRSs without the CSI estimated explicitly.


\subsubsection{Random Phase Algorithm (RPA)}
The RPA has been widely used in the existing literature to optimize the passive reflection coefficients of IRS (see, e.g., \cite{array, scaling, ofdm,double1}) without assuming any knowledge of the CSI. This algorithm is also applicable to our considered integrated IRS-BS system for solving (P1). Specifically, a given number of feasible solutions of $\mv\Theta$ are randomly generated at the BS, where the phase shift of each reflecting element in $\mv\Theta$ is chosen from interval $[0,2\pi)$ uniformly and independently, and then the solution with the maximum objective value of (P1) is selected as the final solution\footnote{In practice, with each realization of $\mv\Theta$, the BS needs to estimate the effective channels from the users to the antenna array of the BS, i.e., $\{\mv h_{k}(\mv\Theta),~k\in\mathcal K\}$, by applying the channel estimation schemes for the conventional multi-antenna BS without integrated IRS.}. Note that this algorithm is in fact used for the initialization of Algorithm \ref{successive} assuming perfect CSI at the BS. The performance of RPA is highly dependent on the phase control of the single-reflection and double-reflection channel components with independent reflection coefficient realizations. However, for the double-reflection channel components,  due to the coupled terms, i.e., $\mv g_{k,j,q}^{n_{j},n_{q}}\vartheta_{j,n_{j}}\vartheta_{q,n_{q}},~j\neq q\in\mathcal J, n_{j}\in\mathcal N_{j}, n_{q}\in\mathcal{N}_{q}$ in (\ref{effective_channel}), independent phase control with randomly generated reflection coefficients for all IRS simultaneously becomes impossible, which may result in the performance loss of RPA. Thus, we use RPA as a benchmark algorithm and propose an improved solution in the following.

\subsubsection{Iterative Random Phase Algorithm (IRPA)}
To further improve the performance of RPA, we propose an IRPA to decouple the design of reflection coefficients for multiple IRSs, where we apply the RPA to optimize the reflection coefficients of one IRS each time only with those of the other $(J-1)$ IRSs being fixed, which operates in an alternate manner over the $J$ IRSs.
We denote $\mv{\tilde\Theta}_{j}\triangleq\{\vartheta_{j,n_j},~n_j\in\mathcal N_j\},~j\in\mathcal J$, and rewrite the effective user-BS channels in (\ref{effective_channel}) as $\mv h_{k}(\{\tilde{\mv\Theta}_{j},~j\in\mathcal J\})$ and the sum-rate of all users in (\ref{sum_rate}) as $R(\{\tilde{\mv\Theta}_{j},~j\in\mathcal J\})$, respectively.

For the given  $\mv{\tilde\Theta}_{q},q\neq j\in\mathcal J$, the effective user $k$-BS channel in (\ref{effective_channel}) w.r.t. the reflection coefficients of IRS $j$, i.e., $\mv{\tilde\Theta}_{j}$, is given by
\begin{align}
\mv h_{k,j}(\mv{\tilde\Theta}_{j})\triangleq\sum_{n_j=1}^{N_j}\mv\gamma_{k,j}^{n_j}\vartheta_{j,n_j}+\mv\beta_{k,j}^{n_j},\label{effective_new}
\end{align}
where 
\begin{align}
\mv\gamma_{k,j}^{n_j}=\mv f_{k,j}^{n_j}+\sum_{q\neq j}\sum_{n_q=1}^{N_{q}}(\mv g_{k,j,q}^{n_j,n_{q}}+\mv g_{k,q,j}^{n_{q},n_j})\vartheta_{q,n_{q}},
\end{align}
and 
\begin{align}
&\mv\beta_{k,j}^{n_j}=\tilde{\mv h}_{k}+\sum_{q\neq j}\sum_{n_q=1}^{N_{q}}\mv f_{k,q}^{n_{q}}\vartheta_{q,n_{q}}\nonumber\\
&+\sum_{q\neq j}\sum_{r\neq q,r\neq j}\sum_{n_{q}=1}^{N_{q}}\sum_{n_{r}=1}^{N_{r}}\mv g_{k,q,r}^{n_{q},n_{r}}\vartheta_{q,n_{q}}\vartheta_{r,n_{r}}
\end{align}
are constants independent to $\tilde{\mv\Theta}_j$. As can be observed, the effective channel vector in (\ref{effective_new}) is linear to the reflection coefficients of IRS $j$ and thus can be effectively controlled. As such, we can employ the RPA to obtain the solution of $\mv{\tilde\Theta}_{j}$ to maximize the objective value of (P1) under given $\{\mv{\tilde\Theta}_{q},~q\neq j\in\mathcal J\}$. In the following, we introduce the proposed IRPA for designing the reflection coefficients of all IRSs without CSI in detail.

Similar to Algorithm \ref{successive}, we first randomly generate $T_{0}\geq 1$ solutions of $\{\tilde{\mv\Theta}_{j},~j\in\mathcal J\}$ by selecting the phase shift of each reflecting element from interval $[0,2\pi)$ independently, and then select the solution achieving the maximum sum-rate as the initial point. With the initialized $\{\tilde{\mv\Theta}_{j},~j\in\mathcal J\}$, we denote its corresponding sum-rate of all users, i.e., $R(\{\tilde{\mv\Theta}_{j},~j\in\mathcal J\})$, as $R_{o}$. Next, we apply the RPA to update the reflection coefficients of IRS $j$ with those of the other IRSs $q\neq j\in\mathcal J$ being fixed sequentially from $j=1$ to $j=J$.\footnote{For the generalized architecture with $\eta>1$ modular antenna arrays, IRS $j,~j\in\mathcal J$ refers to four IRSs surrounding the same antenna module.} Specifically, when applying the RPA to update the reflection coefficients of IRS $j$, we set the number of randomly generated solutions of $\tilde{\mv\Theta}_{j}$ as $T_{j}$ and fix the reflection coefficients of all other IRSs as $\{\tilde{\mv\Theta}_{q}, q\neq j,~q\in\mathcal J\}$. The $t$-th generated solution is denoted as $\tilde{\mv\Theta}_{j,t}$ with the phase shifts of its reflecting elements randomly chosen from interval $[0,2\pi)$, and the corresponding sum-rate of all users is given by $R_{t}(\tilde{\mv\Theta}_{j,t},\{\tilde{\mv\Theta}_{q},~q\neq j\in\mathcal J\})$. Then, the solution which achieves the maximum sum-rate is selected, i.e., $\tilde{\mv\Theta}_{j}^{\star}=\tilde{\mv\Theta}_{j,t_{j}}$ with $t_{j}=\arg\max_{t=1,2,\cdots,T_{j}}R_{t}(\tilde{\mv\Theta}_{j,t},\{\tilde{\mv\Theta}_{q},~q\neq j\in\mathcal J\})$. If the objective value of (P1) is improved, i.e.,  $R(\mv{\tilde\Theta}_{j}^{\star},\{\mv{\tilde\Theta}_{q},~q\neq j\in\mathcal J\})\geq R_{o}$, we update the reflection coefficients of IRS $j$ as $\tilde{\mv\Theta}_{j}=\tilde{\mv\Theta}_{j}^{\star}$ and $R_{o}=R(\mv{\tilde\Theta}_{j}^{\star},\{\mv{\tilde\Theta}_{q},~q\neq j\in\mathcal J\})$; otherwise, we keep the reflection coefficients of IRS $j$ and $R_{o}$ unchanged. After the RPA has been applied to all IRSs, we check whether the increment of the objective value in (P1) is smaller than a given small positive value $\epsilon_{2}$ or the maximum number of iterations $I_{r,\max}$ is reached. If this is the case, the IRPA terminates; otherwise, we repeat the above procedure for updating the reflection coefficients of all IRSs alternately via RPA. We summarize the proposed IRPA to solve (P1) without CSI in Algorithm \ref{ite_rms}. Since we only update $\tilde{\mv\Theta}_{j},~j\in\mathcal J$, when the sum-rate of the users increases compared to the previous iteration, the objective function in (P1) is non-decreasing and Algorithm \ref{ite_rms} is guaranteed to converge. Denote $I_{r}$ as the number of iterations in Algorithm \ref{ite_rms} (i.e., line 6--line 17) with $I_{r}\leq I_{r,\max}$, and thus the total number of randomly generated solutions in Algorithm \ref{ite_rms} is given by $T_{total}=T_{0}+I_{r}\sum_{j=1}^{J}T_{j}$. Note that the computational complexity of Algorithm \ref{ite_rms} mainly lies in the sum-rate calculation for each generated solution of the reflection coefficients, which is given by $\mathcal{O}(M^{2})$. As a result, the overall computational complexity of Algorithm \ref{ite_rms} is $\mathcal{O}(M^{2}T_{total})$.
\begin{algorithm}
\caption{Iterative Random Phase Algorithm (IRPA) Without CSI.}
\label{ite_rms}
\begin{algorithmic}[1]
\REQUIRE {$P$, $\sigma^{2}$, $T_{0}$ and $\{T_{j},~j\in\mathcal J\}$.}
\FOR{$t=1\rightarrow T_{0}$}
\STATE{Randomly generate the phase shift of each reflecting element in $\{\tilde{\mv\Theta}_{j},~j\in\mathcal J\}$ from interval $[0,2\pi)$, which is denoted as $\{\tilde{\mv\Theta}_{j,t},~j\in\mathcal J\}$.}
\STATE{Based on the estimated effective channels from all users, i.e., $\{\mv h_{k}(\{\tilde{\mv\Theta}_{j,t},~j\in\mathcal J\}),~k\in\mathcal K\}$, the BS calculates the sum-rate of all users as $R_{t}(\{\tilde{\mv\Theta}_{j,t},~j\in\mathcal J\})$.}
\ENDFOR
\STATE{Initialize $\tilde{\mv\Theta}_{j}=\tilde{\mv\Theta}_{j,t_{0}},~j\in\mathcal J$, where $t_{0}=\text{arg}\max_{t=1,2,\cdots,T_{0}}R_{t}(\{\tilde{\mv\Theta}_{j,t},~j\in\mathcal J\}$ and the corresponding sum-rate of all users $R(\{\tilde{\mv\Theta}_{j},~j\in\mathcal J\})$ is denoted as $R_{o}$.}
\REPEAT
\FOR{$j=1\rightarrow J$}
\FOR{$t=1\rightarrow T_{j}$}
\STATE{Randomly generate the phase shift of each reflecting element in $\tilde{\mv\Theta}_{j}$ from interval $[0,2\pi)$, which is denoted as $\tilde{\mv\Theta}_{j,t}$.}
\STATE{Based on the estimated effective channels from all users, i.e., $\{\mv h_{k}(\tilde{\mv\Theta}_{j,t},\{\tilde{\mv\Theta}_{q},~q\neq j\in\mathcal J\}),~k\in\mathcal K\}$, the BS calculates the sum-rate of all users as $R_{t}(\tilde{\mv\Theta}_{j,t},\{\tilde{\mv\Theta}_{q},~q\neq j\in\mathcal J\})$. }
\ENDFOR
\STATE{Set $\tilde{\mv\Theta}_{j}^{\star}=\tilde{\mv\Theta}_{j,t_{j}}$ with $t_{j}=\text{arg}\max_{t=1,2,\cdots,T_{j}}R_{t}(\tilde{\mv\Theta}_{j,t},\{\tilde{\mv\Theta}_{q},~q\neq j\in\mathcal J\})$.}
\IF {Objective value of (P1) is improved, i.e., $R(\tilde{\mv\Theta}_{j}^{\star},\{\tilde{\mv\Theta}_{q},~q\neq j\in\mathcal J\})>R_{0}$}
\STATE{Update $\tilde{\mv\Theta}_{j}=\tilde{\mv\Theta}_{j}^{\star}$ and $R_{o}=R(\tilde{\mv\Theta}_{j}^{\star},\{\tilde{\mv\Theta}_{q},~q\neq j\in\mathcal J\})$.}
\ENDIF
\ENDFOR
\UNTIL{the increment of the objective function in (P1) is smaller than $\epsilon_{2}$ or the maximum number of iterations $I_{r,\max}$ is reached. }
\ENSURE {$\{\tilde{\mv\Theta}_{j},~j\in\mathcal J\}$.}
\end{algorithmic}
\end{algorithm}

\section{Numerical Results}\label{numerical_results}
In this section, we provide numerical results to validate the performance of the proposed integrated IRS-BS architecture and passive reflection designs. Unless otherwise specified, the simulation parameters are set as follows. The operating frequency is set as $f_{c}=6~\text{GHz}$, and thus the carrier wavelength is $\lambda=0.05~\text{m}$. The size of the antenna array of the BS is set as $M=16$ with $M_{x}=M_{z}=4$, where the antenna spacing is set as $d_{a}=\lambda/2=0.025~\text{m}$. The number of reflecting elements in each IRS $j\in\mathcal J$ is set to $N_{j}=8$ with $N_{j,1}=1$ and $N_{j,2}=8$, and thus the total number of reflecting elements is $N=\sum_{j=1}^{J}N_{j}=32$. The spacing between two adjacent reflecting elements is set as $d_{I}=\lambda/2=0.025~\text{m}$, the area of each reflecting element is set as $A=(\lambda/2)^{2}$, and the initial phase of the received signals at IRS $j$ relative to the origin $O$ is set to $\varphi_{j}=0,~j\in\mathcal J$, for simplicity. The suspension angle of the antenna radome is set to $\theta_{tilt}=0$, the noise power is set as $\sigma^{2}=-70~\text{dBm}$, and the transmit power at users is set as $P=30~\text{dBm}$. For Algorithm \ref{successive}, we set its stopping threshold as $\epsilon_{1}=10^{-5}$, the maximum number of iterations as $I_{\max}=100$, and the number of random initialization as $T=100$. The number of ground users is set to $K=3$, which are assumed to have quasi-static/slow-fading channels with the antenna array of the BS and all IRSs. For the considered geometric multipath channels between users and the BS, we set $L_{k}=4$ and $\tilde{a}_{k,l}\sim\mathcal{CN}(0,2\times 10^{-12}),~k\in\mathcal K, l\in\mathcal L_{k}$, the elevation angle $\tilde{\theta}_{k,l}$ is randomly generated in interval $[0,\pi/2)$, and the azimuth angle $\tilde{\phi}_{k,l},$ is randomly generated in interval $[0,2\pi)$. Furthermore, we adopt the 3GPP radiation pattern for the antenna elements at the BS \cite{antenna_model}. All results in the simulation are averaged over 100 independent realizations of random channels.

\subsection{Convergence Behaviour of Algorithm \ref{successive}}
\begin{figure}
\centering
\includegraphics[width=8cm]{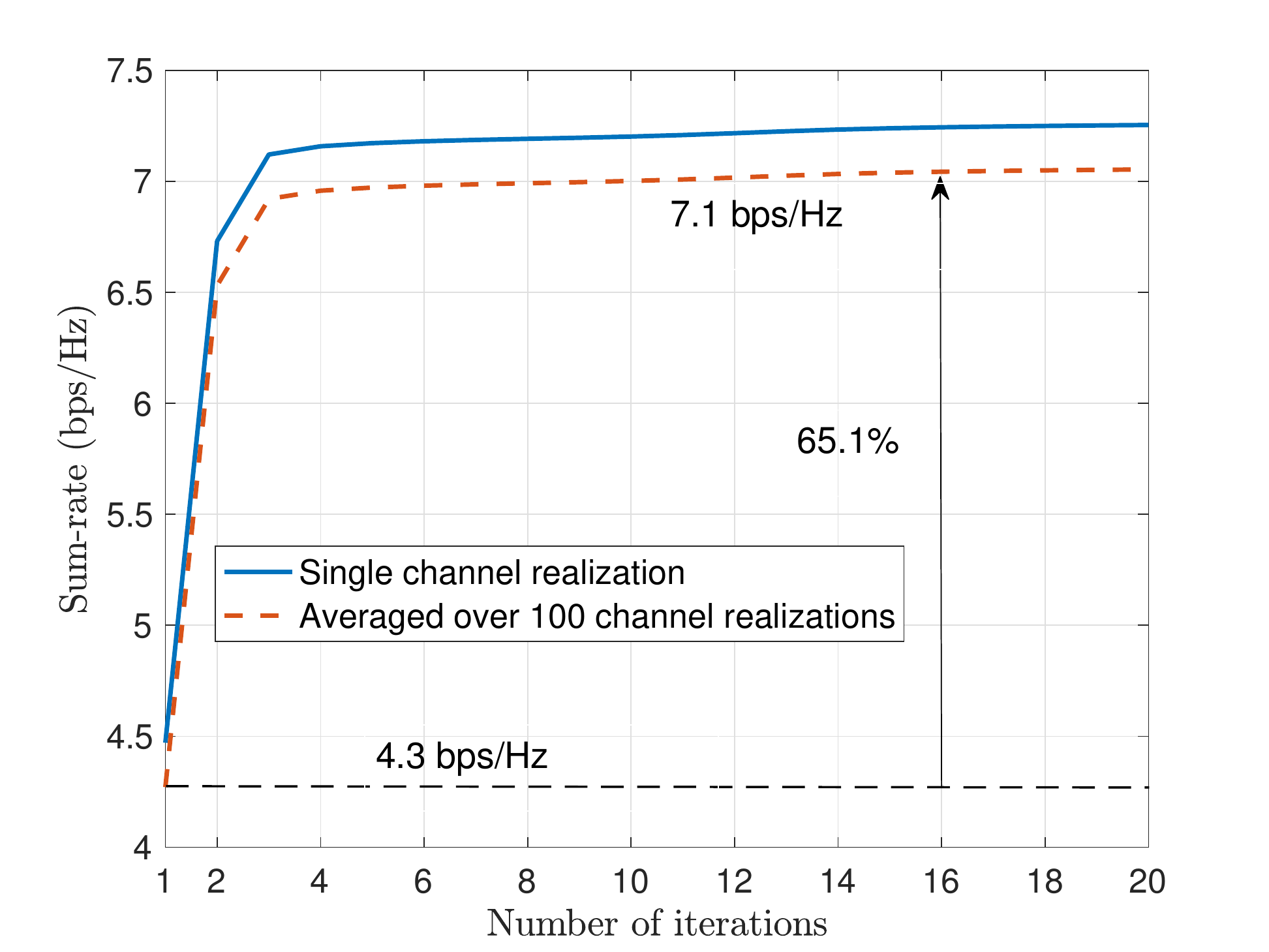}
\vspace{-10pt}\caption{Convergence behaviour of Algorithm \ref{successive}.}\label{convergence}
\vspace{-10pt}\end{figure}
First,  Fig. \ref{convergence} shows the convergence behaviour of Algorithm \ref{successive} under the case of one channel realization as well as that averaged over 100 independent channel realizations. It is observed that the sum-rate monotonically increases for both cases and converges after about 10 iterations. Moreover, for the case of 100 independent channel realizations, the objective value increases from 4.3 bps/Hz to 7.1 bps/Hz during the iterations, which reaps 65.1\% increment on the achievable sum-rate. The results verify the convergence and efficacy of Algorithm \ref{successive} in improving the sum-rate performance of the considered integrated IRS-BS architecture.

\vspace{-10pt}\subsection{Performance Comparison with Conventional Far-Field Channel Model}
Next, we show the superiority of considering element-wise channel model in the proposed architecture by comparing it with the conventional far-field channel model. Note that for the far-field channel model, the UPW propagation is used to model the channel between each IRS and BS antenna array. Then, we apply the widely used method in existing works based on far-field channel model \cite{double1,double2} to model IRS-BS channels and inter-IRS channels (the details are omitted here for brevity). Besides, in these works \cite{double1,double2}, the electromagnetic response for each reflecting element of IRS is assumed to be isotropic for different incident/reflected signals from/towards its reflection half-space, and thus the reflection gain of each IRS defined in (\ref{reflection_gain}) is given by
 \begin{align}\label{reflection_gain_new}
	&G(\hat{\theta}^{A},\hat{\theta}^{D})=\begin{cases}
		2~&\text{if}~\hat{\theta}^{A}\in[0,\frac{\pi}{2})~\text{and}~\hat{\theta}^{D}\in[0,\frac{\pi}{2}),\\
		0,~&\text{otherwise}.
	\end{cases}
\end{align}
Based on the derived far-field channel model, we still solve problem (P1) to design the reflection coefficients of all IRSs, and then substitute the obtained solution to the practical element-wise channel model for performance comparison. 

\begin{figure}
\centering
\includegraphics[width=8cm]{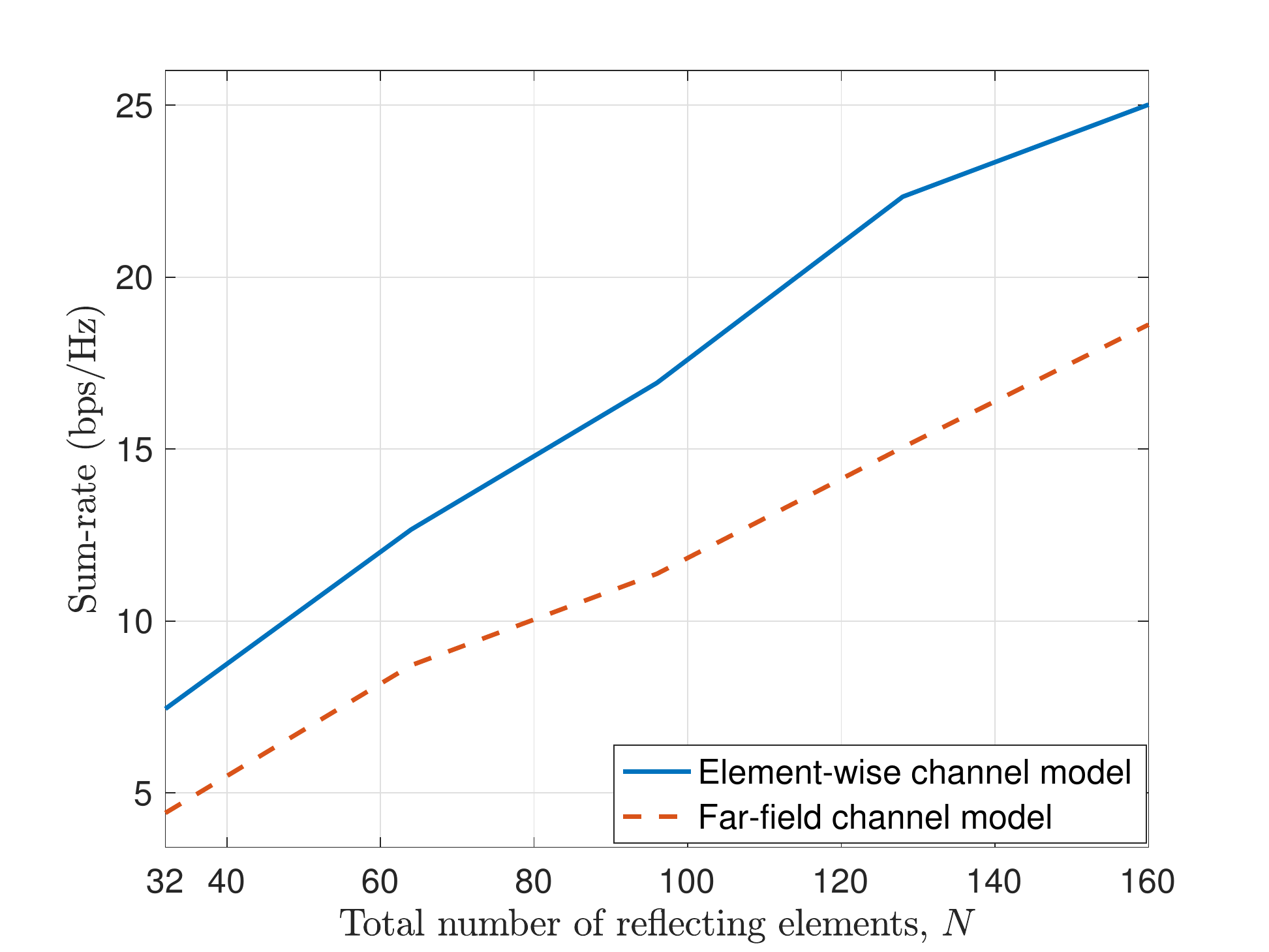}
\vspace{-10pt}\caption{Performance comparison between our proposed element-wise channel model and the conventional far-field channel model.}\label{compare_channel}
\vspace{-10pt}\end{figure}

Fig. \ref{compare_channel} shows the performance achieved by our proposed element-wise channel model and the conventional far-field channel model, respectively, where we increase $N_{j,1}$ from $1$ to $5$ and fix $N_{j,2}=8,~j\in\mathcal J$, and thus the total number of reflecting elements $N$ is increased from $32$ to $160$.\footnote{To reduce the computational complexity, we apply the {\it IRS element grouping} strategy \cite{ofdm,group} when $N\geq 32$ to group the reflecting elements in the same $y$-axis of each IRS into a subsurface, which will employ a common reflection coefficient. Under this strategy, the total number of subsurfaces under all considered $N$ values is fixed as $\tilde{N}=32$.} It is observed that the passive reflection design based on far-field channel model sacrifices the sum-rate performance of users due to the model mismatch, which confirms the necessity of considering the element-wise channel model for the proposed integrated IRS-BS architecture.

\subsection{Performance Comparison between Proposed and Benchmark Algorithms}
Then, we evaluate the efficacy of the proposed Algorithm \ref{successive} with perfect CSI and Algorithm 2 (i.e., IRPA) without CSI by comparing them with the conventional RPA introduced in Section \ref{reflection_design}-C and the following benchmark algorithms.
\begin{itemize}
\item {\bf Low-complexity inexact alternating optimization (LC-IAO) in \cite{LC_IAO}:} In this benchmark algorithm, we assume that the perfect CSI is available at the BS, and then adopt the LC-IAO algorithm in \cite{LC_IAO} to design the reflection coefficients of all IRSs.
\item {\bf Discrete Fourier transform (DFT)-based codebook search (DFT-CS):} In this benchmark algorithm, we consider the case without perfect CSI at the BS and search the optimal codeword in a two-dimensional DFT codebook for designing the IRS passive reflection coefficients. For each IRS $j,~j\in\mathcal J$, the codebook is defined as $\mathcal W_{j}=\{\mv w_{j}|\mv w_{j}=\mv w_{j,1}\otimes \mv w_{j,2}, \mv w_{j,1}\in\mathcal W_{j,1}, \mv w_{j,2}\in\mathcal W_{j,2}\}$, where $\mathcal {W}_{j,1}$ and $\mathcal {W}_{j,2}$ represent the sets of vectors including all the columns of DFT matrices of size $N_{j,1}$ and $N_{j,2}$, respectively. Then, we jointly search the passive reflection coefficients of all IRSs over their given codebooks for achieving the maximum sum-rate $R(\mv\Theta)$ in (\ref{sum_rate}). Thus, the total number of generated solutions for implementing DFT-CS is given by $\prod_{j=1}^{J} N_{j,1}N_{j,2}$.
\end{itemize}

\begin{figure}
\centering
\includegraphics[width=8cm]{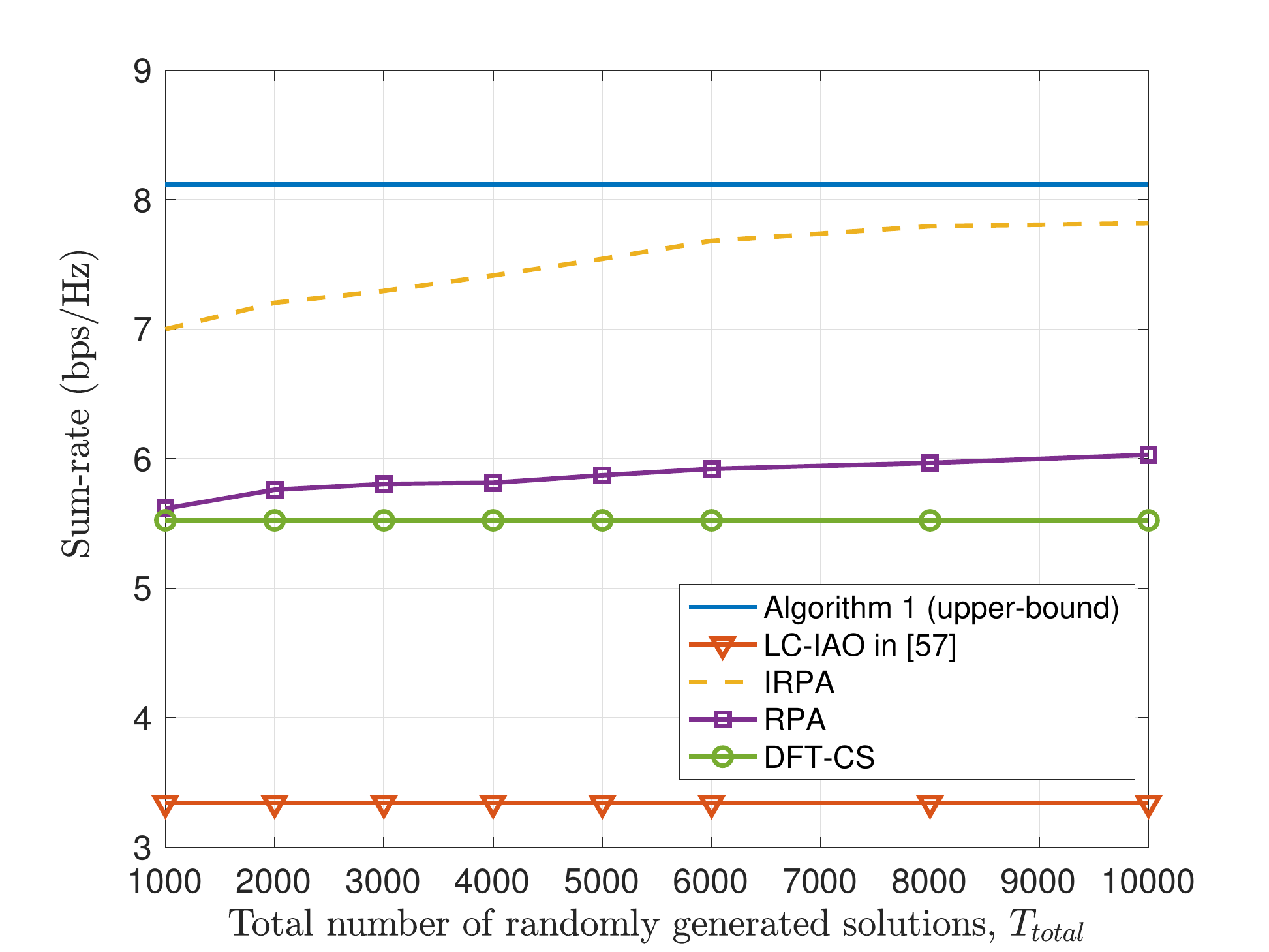}
\vspace{-10pt}\caption{Sum-rate versus the total number of randomly generated solutions, $T_{total}$.}\label{effect_T}
\vspace{-10pt}\end{figure}

Fig. \ref{effect_T} shows the sum-rate of all users versus the total number of randomly generated solutions $T_{total}$. On one hand, for IRPA, we set its stopping threshold as $\epsilon_{2}=10^{-5}$, the maximum number of iterations as $I_{r,\max}=10$, $T_{0}=200$ and $T_{j}=\frac{T_{total}-T_{0}}{4I_{r,\max}},~j\in\mathcal J$. On the other hand, for the RPA, we set its total number of randomly generated solutions for $\{\tilde{\mv\Theta}_{j},~j\in\mathcal J\}$ as $T_{total}$. It is observed that our proposed Algorithm 1 significantly outperforms the LC-IAO proposed in \cite{LC_IAO} in the case with perfect CSI at the BS. It also is observed that the sum-rates achieved by IRPA and RPA increase with $T_{total}$ due to more randomly generated reflection coefficient solutions. As expected, the IRPA outperforms the RPA because the former decouples the design of reflection coefficients for different IRSs and thus more effectively controls the double-reflection channel components. Besides, the IRPA is observed to approach the performance upper-bound achieved by Algorithm \ref{successive} with perfect CSI as $T_{total}$ increases, even without any CSI initially. Furthermore, it is observed that the IRPA achieves better performance than the DFT-CS despite more generated solutions in DFT-CS, i.e., $\prod_{j=1}^{J} N_{j,1}N_{j,2}=4096$ in this example. This indicates that the conventional DFT-CS that performs well in the systems with the UPW propagation adopted for the channels between the IRSs and the antenna array of the BS, as well as only single-reflection channel components present \cite{dft_design1}, is not applicable to our considered integrated IRS-BS architecture, since the far-field condition is not satisfied between the IRSs and the antenna array of the BS, and the more pronounced double-reflection channel components coexist with the single-reflection channel components (see (\ref{effective_channel})) \cite{dft_design2}.

\vspace{-10pt}\subsection{Performance Comparison under Different Channel Setups}
In addition, we aim to evaluate the impact of the single-reflection and double-reflection channel components on the performance of the proposed integrated IRS-BS architecture. Towards this end, we define two benchmark channel setups for performance comparison, where only the single-reflection channel components, i.e., $\mv f_{k,j}^{n_j},~k\in\mathcal K,j\in\mathcal J,n_j\in\mathcal N_j$, are considered (termed as ``single-reflection only''), and only the double-reflection channel components, i.e., $\mv g_{k,j,q}^{n_j,n_q},~k\in\mathcal K, j\neq q\in\mathcal J,n_j\in\mathcal N_j, n_q\in\mathcal N_{q}$, are considered (termed as ``double-reflection only''), respectively. Our proposed channel setup in (\ref{effective_channel}) is termed as ``single- and double- reflection''. We also consider the conventional multi-antenna BS without integrated IRS as a benchmark scheme, which is termed as ``no-IRS''. For all setups employing IRSs, we adopt Algorithm \ref{successive} to design the reflection coefficients of all IRSs by assuming that perfect CSI is available at the BS. 

\begin{figure}
\centering
\includegraphics[width=8cm]{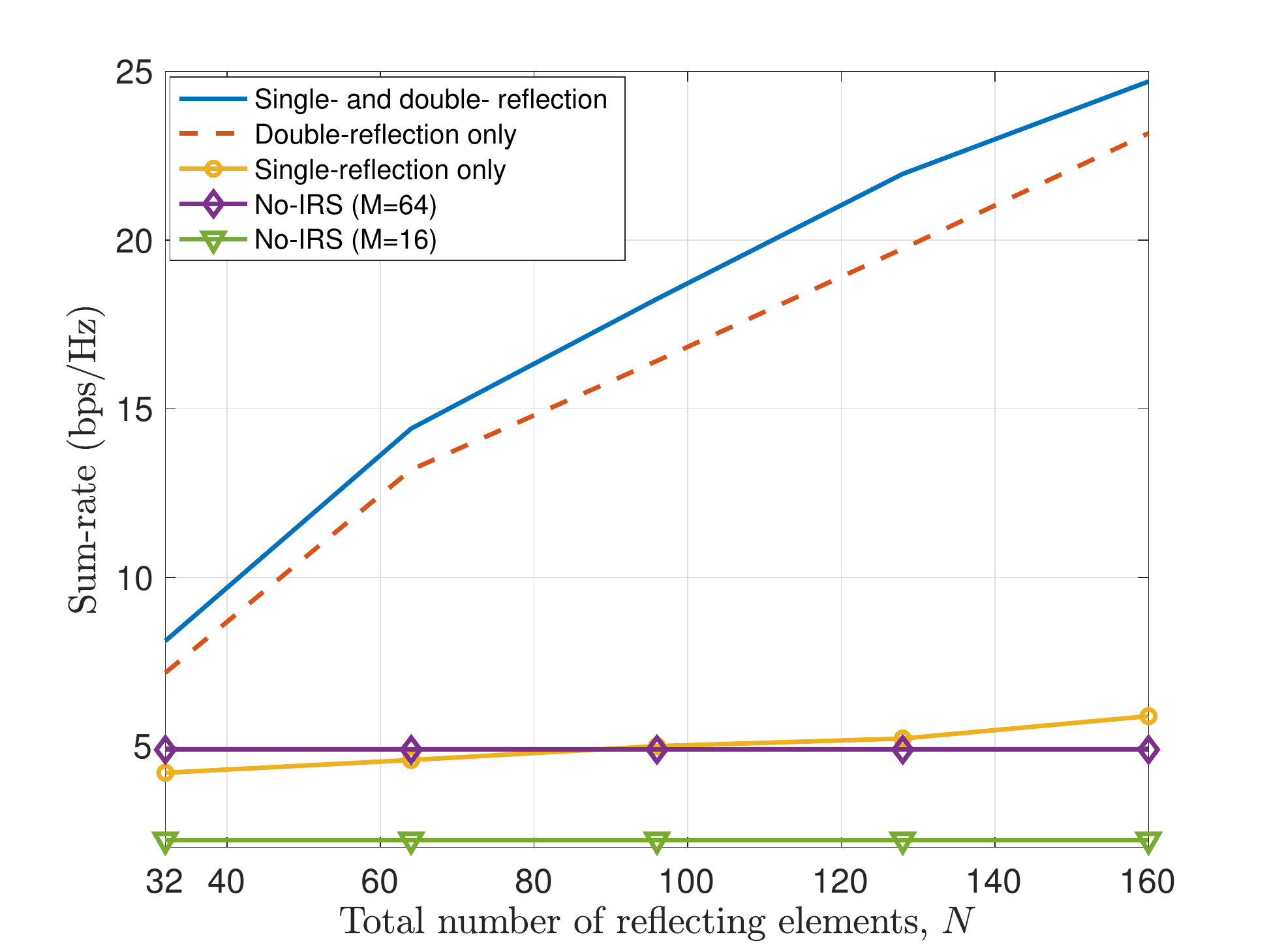}
\vspace{-10pt}\caption{Sum-rate versus the total number of reflecting elements, $N$.}\label{effect_N}
\vspace{-10pt}\end{figure}
Fig. \ref{effect_N} shows the sum-rates achieved by all setups versus the total number of reflecting elements, $N$, where we still increase $N_{j,1}$ from $1$ to $5$ and fix $N_{j,2}=8,~j\in\mathcal J$ to increase the total number of reflecting elements $N$ from $32$ and $160$, and also apply the {\it IRS element grouping strategy} to reduce the complexity. It is observed that the sum-rates achieved by the setups with integrated IRSs monotonically increase with $N$ due to higher passive array reflection gain. It is also observed that the ``double-reflection only'' setup performs closely to ``single- and double- reflection'' setup, and achieves a higher performance gain over the ``single-reflection only'' setup. The performance gap between the ``double-reflection only'' and ``single-reflection only'' setups increases with $N$. This is because in our proposed integrated IRS-BS architecture, the path loss of the double-reflection channel components is not severe due to the short distance between different IRSs. Moreover, the passive array reflection gain provided by the double-reflection channel components is larger than that provided by the single-reflection channel components \cite{double1,double2}, and the number of double-reflection components is much larger than that of single-reflection components, i.e., $\sum_{j=1}^{J}\sum_{q\neq j}^{J}N_{j}N_{q} \gg N$. As a result, the double-reflection channel components have a more dominant impact on the sum-rate performance as compared to the single-reflection channel components. Furthermore, despite a limited number of IRS reflecting elements, the ``single- and double- reflection'' setup significantly outperforms the ``no-IRS'' benchmark scheme employing the same number of antennas (i.e., $M=16$) or even more antennas (i.e., $M=64$). Thus, the proposed integrated IRS-BS architecture is an efficient solution for enhancing the communication rate performance or reducing the number of antennas at the BSs for achieving the same rate performance.

\subsection{Effect of Suspension Angle, $\theta_{tilt}$}
\begin{figure}
\centering
\includegraphics[width=8cm]{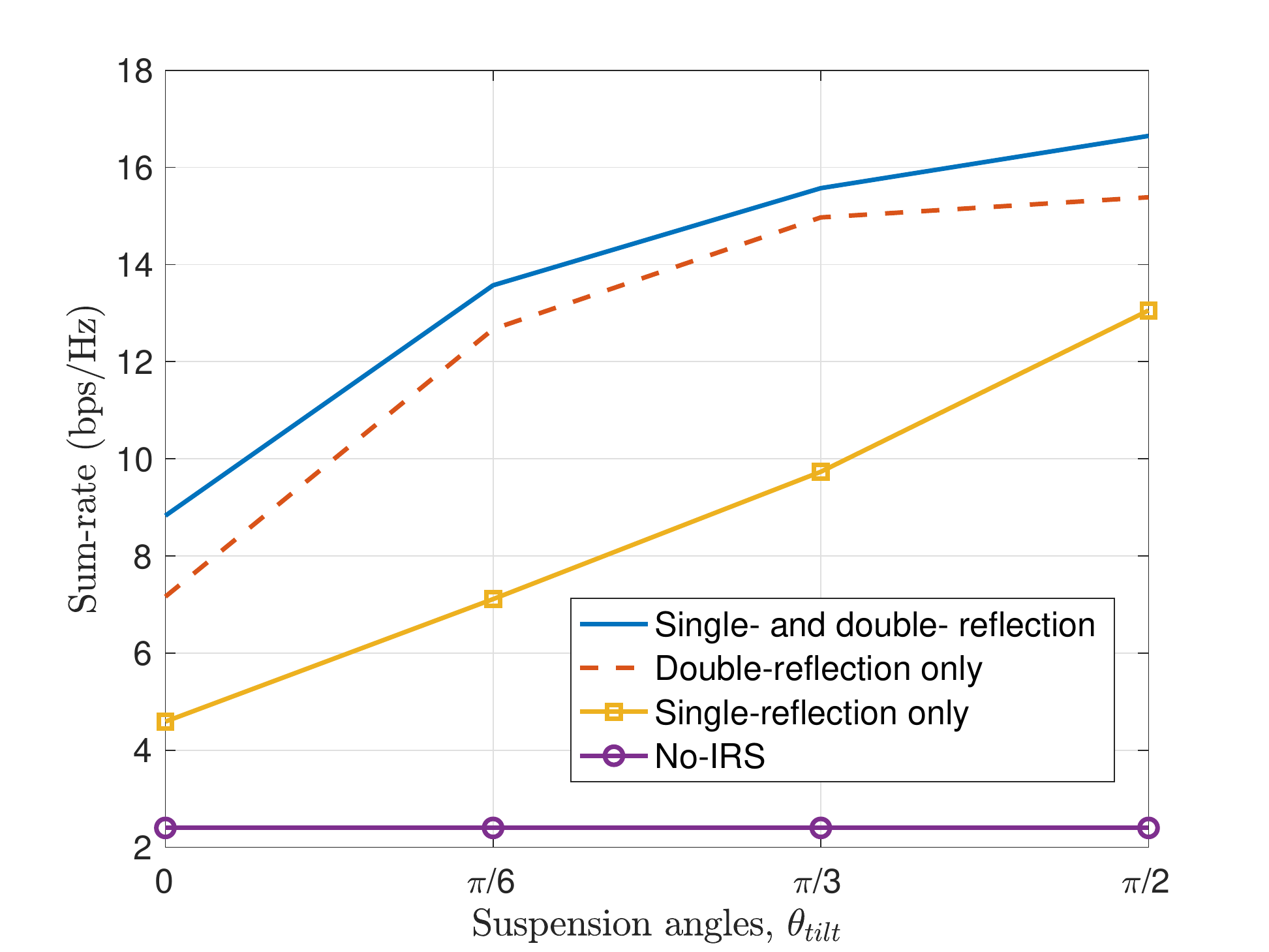}
\vspace{-10pt}\caption{Sum-rates under different suspension angles, $\theta_{tilt}$.}\label{effect_tilt}
\vspace{-10pt}\end{figure}
Furthermore, we investigate the effect of suspension angle $\theta_{tilt}$ on the system performance, where Algorithm \ref{successive} is adopted to design the passive reflection coefficients of all IRSs by assuming perfect CSI at the BS. It is observed from Fig. \ref{effect_tilt} that the sum-rate of users improves by increasing the suspension angle from $\theta_{tilt}=0$ to $\theta_{tilt}=\pi/2$. The reason is that when $\theta_{tilt}=0$ (see Fig. \ref{example} (a)), all users are located outside the reflection half-space of IRS 4, which is deployed at the bottom of the antenna radome (see Fig. \ref{proposed_architecture}). This is more likely to lead to $\hat{\theta}_{k,l,4}>\pi/2,~k\in\mathcal K, l\in\mathcal L_{k}$, and thus IRS 4 cannot provide reflection gains to the single-reflection and double-reflection channel components from {\it all} users to it directly. Specifically, we have $G(\hat{\theta}_{k,l,4},\hat{\theta}_{4,m}^{n_4})=0$ and $G(\hat{\theta}_{k,l,4},\hat{\theta}_{4,j}^{n_4,n_j})=0$, which result in $f_{k,4,m}^{n_4}=0$ and $g_{k,4,j,m}^{n_4,n_j}=0,~m\in\mathcal M,n_{4}\in\mathcal N_{4},j\neq 4\in\mathcal J,n_j\in\mathcal N_j$. By increasing the suspension angle $\theta_{tilt}$, the number of users that can directly receive the reflection gains from IRS 4 via the single-reflection and double-reflection channel components becomes larger, thus enhancing the system rate performance. When the suspension angle is set as $\theta_{tilt}=\pi/2$ (see Fig. \ref{example} (b)), each IRS may provide reflection gains to {\it all} users via all single-reflection and double-reflection channel components, and thus this case achieves the best performance.

\subsection{Performance Comparison under Different Numbers of Antenna Modules for Generalized IRS-BS Architecture with Modular Antenna Arrays}
Finally, we compare the performance of the generalized IRS-BS architecture with modular antenna arrays under different numbers of antenna modules as introduced in Section \ref{system_architecture}-C, where Algorithm \ref{successive} is adopted to design the reflection coefficients of all IRSs. Note that for this generalized IRS-BS architecture,  the LoS channels only exist between each antenna and its surrounding IRSs (modelled as $\xi_{j,m}^{n_j}$ in (\ref{irs_bs})), and between two reflecting elements surrounding the same antenna module (modelled as $\zeta_{j,q}^{n_j,n_q}$ in (\ref{inter_irs})).

\begin{figure}
\centering
\includegraphics[width=8cm]{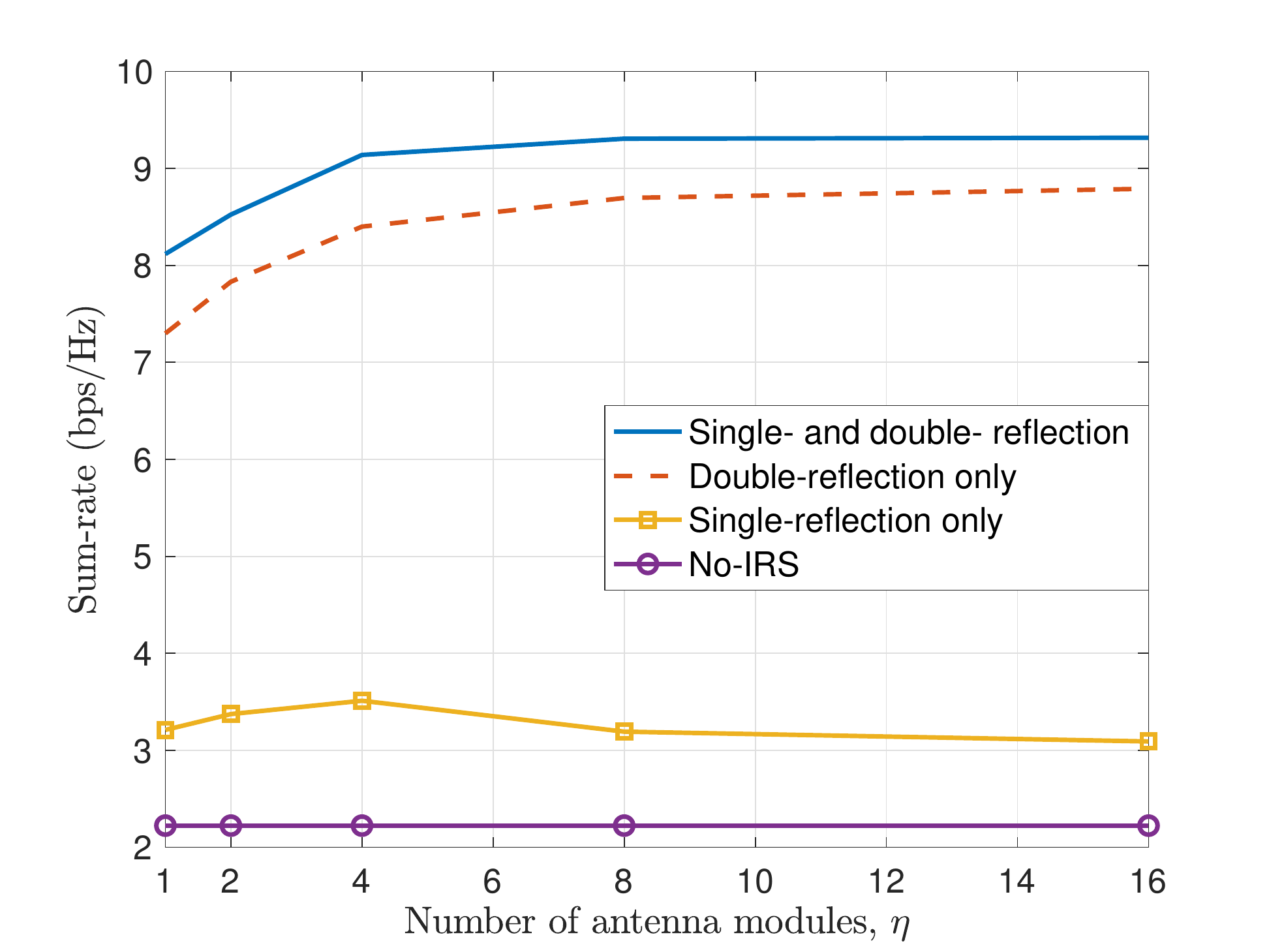}
\vspace{-10pt}\caption{Sum-rate versus the number of antenna modules, $\eta$.}\label{effect_J}
\vspace{-10pt}\end{figure}
Fig. \ref{effect_J} shows the sum-rate versus the number of antenna modules, i.e., $\eta$. It is observed that the performance of the generalized IRS-BS architecture with modular antenna arrays slightly increases with $\eta$ when $\eta\leq 4$ and reaches steady state when $\eta>4$. However, the total number of IRS reflecting elements always increases with $\eta$, which results in an increasing cost at the BS as well as more overhead for channel estimation. Specifically, the sum-rates achieved by the ``single- and double- reflection'' and ``double-reflection only'' setups first increase by 12\% from $\eta=1$ to $\eta=4$ and then become approximately static when $\eta>4$. The sum-rate achieved by ``single-reflection only'' setup first increases by 10\% from $\eta=1$ to $\eta=4$ and then decreases to the same level as that for $\eta=1$. This is because when $\eta\leq 4$, signal propagation distance of the single-reflection and double-reflection channel components decreases as $\eta$ increases, and thus the achievable sum-rate increases. In contrast, when $\eta>4$, the number of single-reflection and double-reflection channel components is reduced, which degrades the sum-rate performance, even though their path loss still decreases as $\eta$ increases. The results in Fig. \ref{effect_J} provide useful guidelines for choosing proper integrated BS-IRS architecture to balance the performance and cost trade-off.

\section{Conclusions}
In this paper, we proposed a new integrated IRS-BS architecture to deploy IRSs with different orientations and an antenna array within the same antenna radome at the BS, and also extended it to a generalized IRS-BS architecture with modular antenna arrays for accommodating more reflecting elements. Considering the ultra-short propagation distance, we proposed an element-wise channel model for  IRS to characterize the direct (without any IRS's reflection) as well as the single and double IRS-reflection channel components between each single-antenna user and the antenna array of the BS, where the UPW was utilized to model the channels from each reflecting element (instead of the whole surface of the IRS) to each antenna at the BS as well as between the reflecting elements of different IRSs. Under the considered channel model, we jointly optimized the reflection coefficients of all IRSs to maximize the uplink sum-rate of the users. By considering two typical cases with/without perfect CSI at the BS, the formulated problem was solved efficiently by exploiting the successive refinement method and IRPA, respectively.

Numerical results demonstrated that the proposed integrated IRS-BS architecture can achieve significant performance gain over the conventional multi-antenna BS without integrated IRS. It was also shown that the developed IRPA can outperform other benchmark algorithms in terms of sum-rate under the case without perfect CSI, and approach the performance upper bound with perfect CSI as the training overhead increases. Moreover, the double-reflection channel components were shown to have a more dominant impact on the achievable sum-rate than the single-reflection channel components due to the short inter-IRS distance and the significantly larger number of double-reflection channel components. Finally, the number of antenna modules in the generalized IRS-BS architecture was shown to have a significant influence on the system rate performance, which should be properly selected in practice to balance the communication performance and implementation cost. It is noteworthy that this paper is an initial study of integrating IRSs into BS/AP, while the results can be extended to other promising directions in future work, such as considering practical IRS phase-shift model \cite{practical_ps}, IRS passive reflection optimization in frequency-selective wideband channels, joint optimization of IRS passive reflection and user transmission scheduling, IRS passive reflection optimization based on statistical CSI to reduce the training overhead, and so on. Besides, under the new element-wise channel model, the theoretical analysis on the proposed integrated IRS-BS architecture, such as deriving its performance scaling laws with increasing number of antennas/IRS elements, is also an interesting problem for future research.

\end{document}